\newcommand{\pcmq}{\mbox{cm$^{-2}$}}
\newcommand{\psec}{\mbox{s$^{-1}$}}

\newcommand{\funit}{\mbox{ph~\pcmq~\psec}}

\newcommand{\eunit}{\mbox{erg~\pcmq~\psec}}

\def\deg{\ensuremath{^\circ}}

\newcommand{\ra}{\mbox{$\alpha_{\rm J2000}$}}
\newcommand{\dec}{\mbox{$\delta_{\rm J2000}$}}
\newcommand{\hi}{\mbox{H\,{\scriptsize I}}}
\newcommand{\hii}{\mbox{H\,{\scriptsize II}}}
\newcommand{\halpha}{\mbox{H$\alpha$}}
\newcommand{\hmol}{\mbox{H$_2$}}
\newcommand{\psmod}{\mbox{PS}}
\newcommand{\gmod}{\mbox{2DG}}

\newcommand{\Msol}{\mbox{$M_{\sun}$}}
\newcommand{\Zsol}{\mbox{$Z_{\sun}$}}

\newcommand{\qunit}{\mbox{ph~s$^{-1}$~sr$^{-1}$~H-atom$^{-1}$}}
\newcommand{\ecut}{\mbox{$E_{\rm c}$}}
\newcommand{\psra}{\mbox{PSR~J0540$-$6919}}
\newcommand{\psrb}{\mbox{PSR~J0537$-$6910}}
\newcommand{\hone}{\mbox{$\mathcal{H}_1$}}
\newcommand{\htwo}{\mbox{$\mathcal{H}_2$}}

\documentclass[longauth]{aa} 

\usepackage{graphicx}
\usepackage{txfonts}
\usepackage{natbib}
\usepackage{lineno}
\bibpunct{(}{)}{;}{a}{}{,} 
\begin{document}

\title{Observations of the Large Magellanic Cloud  with {\em Fermi}}

\author{
A.~A.~Abdo$^{1,2}$ \and
M.~Ackermann$^{3}$ \and 
M.~Ajello$^{3}$ \and
W.~B.~Atwood$^{4}$ \and
L.~Baldini$^{5}$ \and
J.~Ballet$^{6}$ \and
G.~Barbiellini$^{7,8}$ \and
D.~Bastieri$^{9,10}$ \and
B.~M.~Baughman$^{11}$ \and 
K.~Bechtol$^{3}$
% ...
}
\authorrunning{LAT collaboration}

\institute{
\inst{1}~Space Science Division, Naval Research Laboratory, Washington, DC 20375, USA\\
\inst{2}~National Research Council Research Associate, National Academy of Sciences, Washington, DC 20001, USA\\
\inst{3}~W. W. Hansen Experimental Physics Laboratory, Kavli Institute for Particle Astrophysics and Cosmology, Department of Physics and SLAC National Accelerator Laboratory, Stanford University, Stanford, CA 94305, USA\\
\inst{4}~Santa Cruz Institute for Particle Physics, Department of Physics and Department of Astronomy and Astrophysics, University of California at Santa Cruz, Santa Cruz, CA 95064, USA\\
\email{tporter@scipp.ucsc.edu}\\
\inst{5}~Istituto Nazionale di Fisica Nucleare, Sezione di Pisa, I-56127 Pisa, Italy\\
\inst{6}~Laboratoire AIM, CEA-IRFU/CNRS/Universit\'e Paris Diderot, Service d'Astrophysique, CEA Saclay, 91191 Gif sur Yvette, France\\
\inst{7}~Istituto Nazionale di Fisica Nucleare, Sezione di Trieste, I-34127 Trieste, Italy\\
\inst{8}~Dipartimento di Fisica, Universit\`a di Trieste, I-34127 Trieste, Italy\\
\inst{9}~Istituto Nazionale di Fisica Nucleare, Sezione di Padova, I-35131 Padova, Italy\\
\inst{10}~Dipartimento di Fisica ``G. Galilei", Universit\`a di Padova, I-35131 Padova, Italy\\
\inst{11}~Department of Physics, Center for Cosmology and Astro-Particle Physics, The Ohio State University, Columbus, OH 43210, USA\\
\inst{12}~Istituto Nazionale di Fisica Nucleare, Sezione di Perugia, I-06123 Perugia, Italy\\
\inst{13}~Dipartimento di Fisica, Universit\`a degli Studi di Perugia, I-06123 Perugia, Italy\\
\inst{14}~Dipartimento di Fisica ``M. Merlin" dell'Universit\`a e del Politecnico di Bari, I-70126 Bari, Italy\\
\inst{15}~Istituto Nazionale di Fisica Nucleare, Sezione di Bari, 70126 Bari, Italy\\
\inst{16}~Laboratoire Leprince-Ringuet, \'Ecole polytechnique, CNRS/IN2P3, Palaiseau, France\\
\inst{17}~Department of Physics, University of Washington, Seattle, WA 98195-1560, USA\\
\inst{18}~INAF-Istituto di Astrofisica Spaziale e Fisica Cosmica, I-20133 Milano, Italy\\
\inst{19}~NASA Goddard Space Flight Center, Greenbelt, MD 20771, USA\\
\inst{20}~Center for Research and Exploration in Space Science and Technology (CRESST), NASA Goddard Space Flight Center, Greenbelt, MD 20771, USA\\
\inst{21}~University of Maryland, Baltimore County, Baltimore, MD 21250, USA\\
\inst{22}~George Mason University, Fairfax, VA 22030, USA\\
\inst{23}~Laboratoire de Physique Th\'eorique et Astroparticules, Universit\'e Montpellier 2, CNRS/IN2P3, Montpellier, France\\
\inst{24}~Department of Physics and Astronomy, Sonoma State University, Rohnert Park, CA 94928-3609, USA\\
\inst{25}~Department of Physics, Stockholm University, AlbaNova, SE-106 91 Stockholm, Sweden\\
\inst{26}~The Oskar Klein Centre for Cosmoparticle Physics, AlbaNova, SE-106 91 Stockholm, Sweden\\
\inst{27}~Royal Swedish Academy of Sciences Research Fellow, funded by a grant from the K. A. Wallenberg Foundation\\
\inst{28}~Agenzia Spaziale Italiana (ASI) Science Data Center, I-00044 Frascati (Roma), Italy\\
\inst{29}~Dipartimento di Fisica, Universit\`a di Udine and Istituto Nazionale di Fisica Nucleare, Sezione di Trieste, Gruppo Collegato di Udine, I-33100 Udine, Italy\\
\inst{30}~Universit\'e de Bordeaux, Centre d'\'Etudes Nucl\'eaires Bordeaux Gradignan, UMR 5797, Gradignan, 33175, France\\
\inst{31}~CNRS/IN2P3, Centre d'\'Etudes Nucl\'eaires Bordeaux Gradignan, UMR 5797, Gradignan, 33175, France\\
\inst{32}~Department of Physical Sciences, Hiroshima University, Higashi-Hiroshima, Hiroshima 739-8526, Japan\\
\inst{33}~University of Maryland, College Park, MD 20742, USA\\
\inst{34}~Istituto Nazionale di Fisica Nucleare, Sezione di Trieste, and Universit\`a di Trieste, I-34127 Trieste, Italy\\
\inst{35}~Columbia Astrophysics Laboratory, Columbia University, New York, NY 10027, USA\\
\inst{36}~University of Alabama in Huntsville, Huntsville, AL 35899, USA\\
\inst{37}~Department of Physics, Royal Institute of Technology (KTH), AlbaNova, SE-106 91 Stockholm, Sweden\\
\inst{38}~Department of Physics, Tokyo Institute of Technology, Meguro City, Tokyo 152-8551, Japan\\
\inst{39}~Waseda University, 1-104 Totsukamachi, Shinjuku-ku, Tokyo, 169-8050, Japan\\
\inst{40}~Cosmic Radiation Laboratory, Institute of Physical and Chemical Research (RIKEN), Wako, Saitama 351-0198, Japan\\
\inst{41}~Centre d'\'Etude Spatiale des Rayonnements, CNRS/UPS, BP 44346, F-31028 Toulouse Cedex 4, France\\
\email{pierre.jean@cesr.fr}\\
\email{jurgen.knodlseder@cesr.fr}\\
\inst{42}~Max-Planck Institut f\"ur extraterrestrische Physik, 85748 Garching, Germany\\
\inst{43}~Istituto Nazionale di Fisica Nucleare, Sezione di Roma ``Tor Vergata", I-00133 Roma, Italy\\
\inst{44}~Department of Physics and Astronomy, University of Denver, Denver, CO 80208, USA\\
\inst{45}~Institut f\"ur Astro- und Teilchenphysik and Institut f\"ur Theoretische Physik, Leopold-Franzens-Universit\"at Innsbruck, A-6020 Innsbruck, Austria\\
\inst{46}~Institut de Ciencies de l'Espai (IEEC-CSIC), Campus UAB, 08193 Barcelona, Spain\\
\inst{47}~Space Sciences Division, NASA Ames Research Center, Moffett Field, CA 94035-1000, USA\\
\inst{48}~NYCB Real-Time Computing Inc., Lattingtown, NY 11560-1025, USA\\
\inst{49}~Department of Chemistry and Physics, Purdue University Calumet, Hammond, IN 46323-2094, USA\\
\inst{50}~Instituci\'o Catalana de Recerca i Estudis Avan\c{c}ats, Barcelona, Spain\\
\inst{51}~Consorzio Interuniversitario per la Fisica Spaziale (CIFS), I-10133 Torino, Italy\\
\inst{52}~Institute of Space and Astronautical Science, JAXA, 3-1-1 Yoshinodai, Sagamihara, Kanagawa 229-8510, Japan\\
\inst{53}~North-West University, Potchefstroom Campus, Potchefstroom 2520, South Africa\\
\inst{54}~Dipartimento di Fisica, Universit\`a di Roma ``Tor Vergata", I-00133 Roma, Italy\\
\inst{55}~Australia Telescope National Facility, CSIRO, Epping NSW 1710, Australia\\
\inst{56}~School of Pure and Applied Natural Sciences, University of Kalmar, SE-391 82 Kalmar, Sweden
}

\date{Received October xx, 2009; January xx, 2010}

\abstract
%
{The Large Magellanic Cloud (LMC) is to date the only normal external galaxy that
has been detected in high-energy gamma rays.
High-energy gamma rays trace particle acceleration processes and gamma-ray observations
allow the nature and sites of acceleration to be studied.}
%
{We characterise the distribution and sources of cosmic rays in the LMC from analysis of 
gamma-ray observations.}
%
{We analyse 11 months of continuous sky-survey observations obtained with the Large Area
Telescope aboard the Fermi Gamma-Ray Space Telescope and compare it to tracers of the 
interstellar medium and models of the gamma-ray sources in the LMC.}
%
{The LMC is detected at 33$\sigma$ significance.
The integrated $>$100~MeV photon flux of the LMC amounts to
$(2.6 \pm 0.2) \times 10^{-7}$ \funit\
which corresponds to an energy flux of
$(1.6 \pm 0.1) \times 10^{-10}$ \eunit, 
with additional systematic uncertainties of $\la16\%$.
The analysis reveals the massive star forming region 30~Doradus as a bright source
of gamma-ray emission in the LMC in addition to fainter emission regions found in the
northern part of the galaxy.
The gamma-ray emission from the LMC shows very little correlation with gas density
and is rather correlated to tracers of massive star forming regions.
The close confinement of gamma-ray emission to star forming regions suggests
a relatively short GeV cosmic-ray proton diffusion length.
}
%
{
The close correlation between cosmic-ray density and massive star tracers supports the idea 
that cosmic rays are accelerated in massive star forming regions as a result of the large 
amounts of kinetic energy that are input by the stellar winds and supernova explosions of 
massive stars into the interstellar medium.
}

\keywords{Acceleration of particles -- cosmic rays -- Magellanic Clouds -- Gamma rays: observations}

\maketitle

\section{Introduction}

Since the early days of high-energy gamma-ray astronomy, it has been clear that the
gamma-ray flux received on Earth is dominated by emission from the Galactic disk
\citep{clark68}.
This emission is believed to arise from cosmic-ray interactions with the interstellar medium, 
which at gamma-ray energies $\ga$~100~MeV are dominated by the decay of $\pi^0$ 
produced in collisions between cosmic-ray nuclei and the interstellar medium \citep{pollack63}.
Further contributions are from cosmic-ray electrons undergoing inverse Compton scattering
off interstellar soft photons and Bremsstrahlung losses within the interstellar medium.
Gamma-ray observations thus have the potential to map cosmic-ray acceleration sites in our 
Galaxy, which may ultimately help to identify the sources of cosmic-ray acceleration.

Nearby galaxies have some advantages as targets for studies of cosmic-ray physics and have
the advantage of being viewed from outside, and so line of sight confusion, which complicates 
studies of emission from the Galactic disk, is diminished.
This advantage is, however, somewhat offset by the limitations by the angular resolution 
and sensitivity of the instrument.
The Large Magellanic Cloud (LMC) is thus an excellent target for studying the link between 
cosmic-ray acceleration and gamma-ray emission since 
the galaxy is nearby \citep[$D\approx50$~kpc;][]{matsunaga09,pietrzynski09},
has a large angular extent of $\sim8\deg$,
and is seen at a small inclination angle of $i\approx20\deg-35\deg$
\citep{kim98,vandermarel06},
which avoids source confusion.
In addition, the LMC is relatively active, housing many supernova remnants, bubbles
and superbubbles, and massive star-forming regions that are all potential sites
of cosmic-ray acceleration \citep{cesarsky83,biermann04,binns07}.

The EGRET telescope aboard the Compton Gamma-Ray Observatory 
({\em CGRO}, 1991--2000) was the first to detect the LMC \citep{sreekumar92}, 
which has remained the only normal galaxy besides our Milky Way that has been seen 
in high-energy gamma rays.
Due to EGRET's limited angular resolution and limited sensitivity, details of the
spatial structure of the gamma-ray emission could not be resolved, yet the 
observations showed some evidence that the spatial distribution is consistent
with the morphology of radio emission.
The observations also allowed determining the integral gamma-ray flux from
the LMC, which was shown to be consistent with expectations based on a model
using the principles of dynamic balance and containment \citep{fichtel91}.
From the agreement with this model, \citet{sreekumar92} conclude that the level
of cosmic rays in the LMC is comparable to that in our Galaxy.

The Large Area Telescope (LAT) aboard {\em Fermi} is now for the first time providing
the capabilities to go well beyond the study of the integrated gamma-ray flux from the 
LMC \citep{digel00,weidenspointner07}.
Thanks to its excellent sensitivity and good angular resolution, we are able to spatially
resolve the structure of the gamma-ray emission, allowing us to provide 
a detailed mapping of the cosmic-ray density in the galaxy.
In the spectral domain, our data allow for detailed comparisons with cosmic-ray
interaction models and can be used to search for spectral variations over the galaxy.
In the timing domain, we are able to assess contributions from flaring sources and
can search for pulsations from energetic pulsars in the galaxy that may
contribute to the total emission \citep{harding81,hartmann93}.

In this paper we present our first in-depth analysis of the LMC galaxy based on 11 months
of continuous sky survey observations performed with {\em Fermi}/LAT.
We put particular emphasis on determining the spatial distribution of the
gamma-ray emission, which, as we will show, reveals the distribution of cosmic rays
in the galaxy.
We also search pulsations from energetic pulsars that may contribute to a 
non-negligible level to the overall emission.

\section{Observations}
\label{sec:observation}

\subsection{Data preparation}

The characteristics and performance of the LAT aboard {\em Fermi}
are described in detail by \citet{atwood09}.
The data used in this work amount to 274.3 days of continuous sky survey observations
over the period
August 8 2008 -- July 9 2009
during which a total exposure of 
$\sim2.5 \times 10^{10}$~cm$^2$~s (at 1 GeV)
has been obtained for the LMC.
Events satisfying the standard low-background event
selection \citep[``Diffuse'' events;][]{atwood09} and coming from zenith angles
$<105\deg$ (to greatly reduce the contribution by Earth albedo gamma rays)
were used.
To further reduce the effect of Earth albedo backgrounds, the time intervals
when the Earth was appreciably within the field of view (specifically, when
the centre of the field of view was more than $47\deg$ from the zenith)
were excluded from this analysis.
Furthermore,  time intervals when the spacecraft was within the South Atlantic Anomaly 
were also excluded.
We further restricted the analysis to photon energies above 200~MeV; below this energy,
the effective area in the ``Diffuse class" is relatively small and strongly dependent on 
energy.

For the analysis we selected all events within a rectangular region-of-interest
(ROI) of size $20\deg \times 20\deg$ centred on 
$(\ra, \dec)=(05^{\rm h}17^{\rm m}36^{\rm s}, -69\deg01'48")$\footnote{
 corresponding to $(l,b)=(279.73\deg,-33.54\deg)$}
and aligned in equatorial coordinates.
All analysis was performed using the LAT Science Tools package, which is available 
from the Fermi Science Support Center, using P6\_V3 post-launch instrument response 
functions (IRFs).
These take into account pile-up and accidental coincidence effects in the
detector subsystems that were not considered in the definition of the pre-launch
IRFs.

\subsection{Data selection and background modelling}
\label{sec:backgroundmodel}

At the Galactic latitude of the LMC ($b \approx -33\deg$), the gamma-ray background is
a combination of extragalactic and Galactic diffuse emissions and
some residual instrumental background.
The extragalactic component comprises resolved sources, which often can be associated 
with known blazars \citep{abdo09b}, and a diffuse component, which is attributed
to unresolved sources and eventually intrinsically diffuse processes 
\citep{strong04a,dermer07}.
The latest LAT collaboration-internal source list that has been derived from 11 months of
survey data (comparable to the data volume used in this analysis) contains 542 sources for
latitudes $|b| \ge 30\deg$, corresponding to a source density of $86$ sources sr$^{-1}$.
Our ROI covers a solid angle of 0.12~sr, so we expect about 10 resolved background sources 
in our field.
Among those, 1-2 should spatially overlap with the LMC if we assume the galaxy's diameter
is between $8\deg$ and $10\deg$.

\begin{table}[!t]
\footnotesize
\caption{Point sources included in the background model.
\label{tab:crates}
}
\begin{center}
\begin{tabular}{lrrr}
\hline\hline
Name &  \multicolumn{1}{c}{TS} & \multicolumn{1}{c}{$\ra$} & \multicolumn{1}{c}{$\dec$} \\
\hline
CRATES J050755$-$610441 & 118.6 & $05^{\rm h}07^{\rm m}55^{\rm s}$ & $-61\deg04'43"$ \\
CRATES J051643$-$620706 & 269.3 & $05^{\rm h}16^{\rm m}45^{\rm s}$ & $-62\deg07'05"$ \\
CRATES J055842$-$745904 & 53.5 & $05^{\rm h}58^{\rm m}46^{\rm s}$ & $-74\deg59'05"$ \\
CRATES J060106$-$703606 & 52.5 & $06^{\rm h}01^{\rm m}11^{\rm s}$ & $-70\deg36'09"$ \\
CRATES J063542$-$751615 & 90.0 & $06^{\rm h}35^{\rm m}46^{\rm s}$ & $-75\deg16'16"$ \\
CRATES J070027$-$661041 & 123.4 & $07^{\rm h}00^{\rm m}31^{\rm s}$ & $-66\deg10'45"$ \\
\hline
\end{tabular}
\end{center}
Note to the table: TS is a measure of the detection significance of the source (cf. \ref{sec:geomodels}). 
\end{table}

\begin{figure*}
\centering
\includegraphics[width=9.1cm]{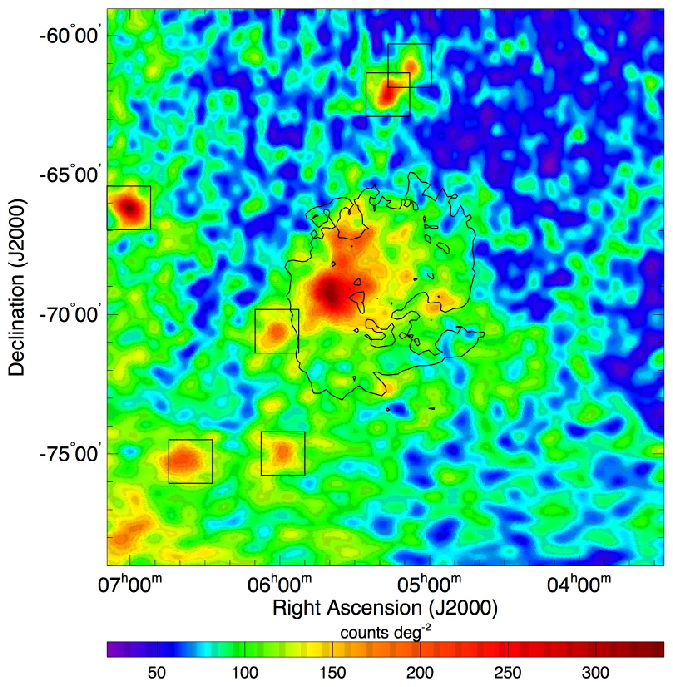}
\includegraphics[width=9.1cm]{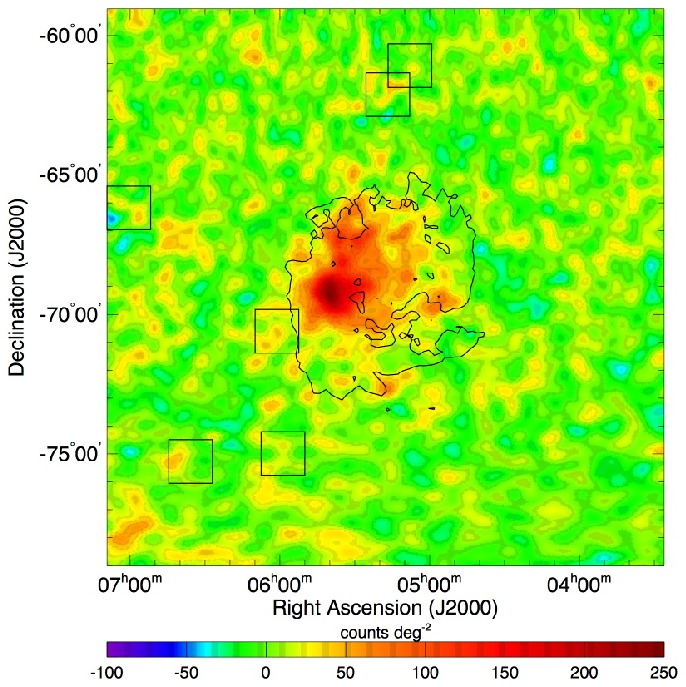}
\caption{
Gaussian kernel ($\sigma=0.2\deg$) smoothed counts maps of the region of interest (ROI) 
in a true local projection before (left) and after subtraction of the background model (right) 
for the energy range 200 MeV -- 20 GeV and for a pixel size of $0.1\deg \times 0.1\deg$.
Overlaid is the N(\hi) contour of $1\times10^{21}$ H~cm$^{-2}$ of the LMC to 
indicate the extent and shape of the galaxy.
The boxes show the locations of the 6 point sources that have been included
in the background model.
The right panel has a true dynamic range from $-46$ to $+248$ counts~deg$^{-2}$
that has been expanded for display to cover the full dynamic range of the residuals that 
are shown in Fig.~\ref{fig:resmaps}.
\label{fig:image}
}
\end{figure*}

Within the ROI but outside the boundaries of the LMC we found a total of 6 significant point 
sources, compatible with expectations (cf.~Table~\ref{tab:crates}).
For all of them we found counterparts in the CRATES catalogue of flat-spectrum radio sources
\citep{healey07} using the procedure described in \citet{abdo09a},
making them good candidates for being background blazars.
Searching for background sources within the LMC boundaries is more difficult due to 
possible confusion with local emission maxima of the galaxy's diffuse emission.
Background blazars may however reveal themselves by their flaring activity, hence
we searched for any excess emission above the time averaged level on a monthly
basis (cf.~section~\ref{sec:variability}).
We found evidence for a flaring source during month 4 of our dataset (MJD 54777.8 - 54808.2)
near 30~Doradus, and thus we excluded the data within this time interval from our analysis.
This results in a dataset that corresponds to 248.7 days of continuous sky survey observations
during which a total exposure of 
$\sim2.3 \times 10^{10}$~cm$^2$~s (at 1 GeV)
was obtained for the LMC.

We then modelled background gamma-ray emission within the ROI using components for the 
diffuse Galactic and the extragalactic and residual instrumental backgrounds and the 
6 blazars.
The Galactic component was based on the LAT standard diffuse background model
{\tt gll\_iem\_v02}\footnote{
The model can be downloaded from
http://fermi.gsfc.nasa.gov/ssc/data/access/lat/BackgroundModels.html.
}
for which we kept the overall normalisation as a free parameter.
The extragalactic and residual instrumental backgrounds were combined into a single
component which has been taken as being isotropic.
The spectrum of this component was determined by fitting an isotropic component
together with a model of the Galactic diffuse emission and point sources to the data.
Also here we left the overall normalisation of the component as a free parameter.
The 6 background blazars were modelled as point sources with power-law spectral shapes.
The positions of the blazars were fixed to those given in the CRATES catalogue
\citep{healey07} and are given in Table \ref{tab:crates}.
The flux and spectral power-law index of each source were left as free parameters of our 
background model and their values were determined from likelihood analysis.

\subsection{Spatial distribution}
\label{sec:spatial}

\subsubsection{Counts map}

To investigate the spatial distribution of gamma-ray emission toward the LMC we first binned all 
photons into a counts map of size $20\deg \times 20\deg$ centred on 
$(\ra, \dec)=(05^{\rm h}17^{\rm m}36^{\rm s}, -69\deg01'48")$
and aligned in equatorial coordinates.
Figure~\ref{fig:image} shows the counts map before (left panel) and after (right panel)
subtraction of the background model.

\begin{figure}
\centering
\includegraphics[width=9.1cm]{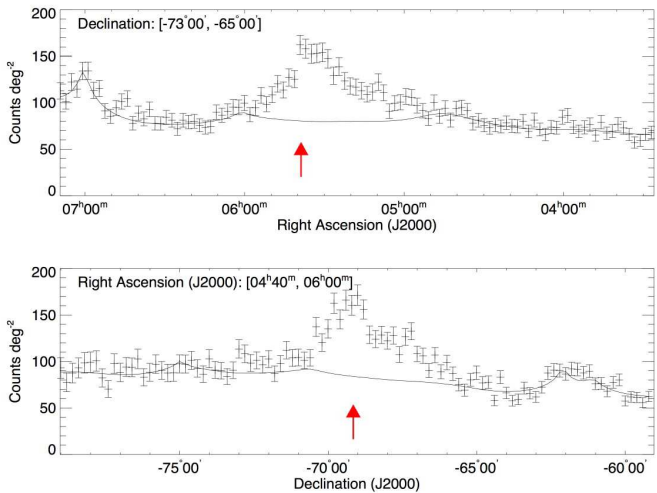}
\caption{
Profiles of the observed number of counts along Right Ascension (top) and Declination (bottom) 
of the LMC region for the energy range 200 MeV -- 20 GeV.
The solid lines show the corresponding profiles of the background model.
The red arrow indicates the location of 30~Dor.
\label{fig:profile}
}
\end{figure}

The background subtracted map confirms that diffuse Galactic and isotropic backgrounds as 
well as the 6 background blazars are properly removed by our treatment.
The only remaining feature is extended emission that is spatially confined to within the LMC
boundaries which we trace by the iso column density contour 
$N_{\rm H} = 1\times10^{21}$ H~cm$^{-2}$
of neutral hydrogen in the LMC \citep{kim05}.
The total number of excess 200 MeV -- 20 GeV photons above the background in the LMC 
area\footnote{
  We use a square region $5.5\deg \times 5.5\deg$ centred on
  $(\ra, \dec)=(05^{\rm h}30^{\rm m}, -68\deg30')$
  to extract these numbers.}
amounts to $\sim1550$ counts whereas the background in the same area amounts to 
$\sim2440$ counts.
With these statistics, the extended gamma-ray emission from the LMC can be
resolved into several components.
The brightest emission feature is located near
$(\ra, \dec) \approx (05^{\rm h}40^{\rm m}, -69\deg15')$,
which is close to the massive star-forming region 30~Doradus (30~Dor)
that houses the two Crab-like pulsars 
\psra\ and \psrb\ \citep{seward84,marschall98}.
Excess gamma-ray emission is also seen toward the north and the west of 30~Dor.
These bright regions are embedded into a more extended and diffuse glow that
covers an area of approximately $5\deg \times 5\deg$.
To further illustrate the emission structure we show profiles of the counts map
in Fig.~\ref{fig:profile} along \ra\ (top panel) and \dec\ (bottom panel) and we provide
an adaptively smoothed image of the 30~Dor region in Fig.~\ref{fig:30Dor}.

\subsubsection{Geometrical models}
\label{sec:geomodels}

Out next step was to assess the spatial distribution of the LMC emission using simple
parametrised geometrical models of the gamma-ray intensity distribution.
We assumed power-law spectral distributions for all models and kept the 
total flux and power law index as free parameters.
We adjusted the spatial and spectral parameters of the models using a binned maximum 
likelihood analysis with spatial bins of $0.1\deg \times 0.1\deg$ and 60 logarithmically 
spaced energy bins covering the energy range 200 MeV -- 20 GeV.
For each model we computed the so-called {\em Test Statistic} (TS), which is defined as 
twice the difference between 
the log-likelihood $\mathcal{L}_1$ that is obtained by fitting 
the model on top of the background model to the data, and
the log-likelihood $\mathcal{L}_0$ that is obtained by fitting the background model only,
i.e. ${\rm TS} = 2(\mathcal{L}_1 - \mathcal{L}_0)$.
Under the hypothesis that the background model satisfactorily explains our data, TS follows 
the $\chi^2_p$ distribution with $p$ degrees of freedom, where $p$ is the number of additional 
free parameters in the maximisation of $\mathcal{L}_1$ with respect to those used in the 
maximisation of $\mathcal{L}_0$ \citep{cash79}.
For example, for a single source on top of the background model with free position, flux and 
spectral power law index we have $p=4$.

First, we examined whether the gamma-ray emission from the LMC can be explained with a
combination of individual point sources.
For this purpose we added successive point sources to our model and optimised their
locations, fluxes and spectral indices by maximising the likelihood of the model.
We stopped this procedure once the TS improvement after adding a further point source
dropped below 25.
This happened after we added a 6th point source to our model which resulted in a 
TS improvement of only 21.9, corresponding to a detection significance of $3.7\sigma$
for this source.
Table \ref{tab:psmodel} provides the maximum likelihood positions and fluxes for the 5
significant point sources, and gives also the TS improvement ($\Delta$TS) that is achieved 
in each of the successive steps.
Adding up $\Delta$TS for all sources provides a total TS of 1089.3 for this model, corresponding
to a detection significance of $31.5\sigma$ ($p=20$).
We refer to the point source model as \psmod.

\begin{figure}[!t]
\centering
\includegraphics[width=9.1cm]{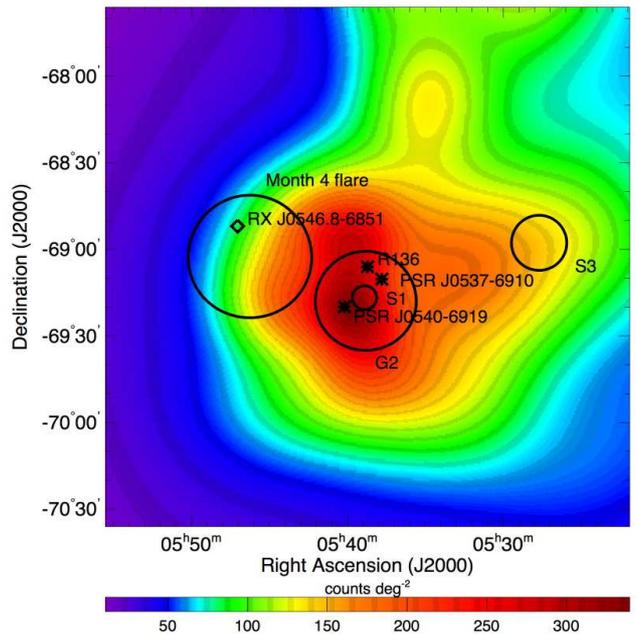}
\caption{
Zoom into a $3\deg \times 3\deg$ large region of the background subtracted counts map 
around the central star cluster R~136 in 30~Dor.
The counts map has been adaptively smoothed \citep{ebeling06} with a signal-to-noise 
ratio of 5 in order to reveal significant structures at all possible scales
while suppressing the noise that arises from photon counting statistics.
Stars show the locations of R~136 and of the pulsars \psra\ and \psrb.
The circles show the 95\% containment radius of sources S1, S3, and G2
(see text).
For G2 we added the Gaussian $\sigma$ of the source to the error radius.
We also show the 95\% containment radius of a flaring source that appeared 
during month 4 of the analysed data set.
The diamond shows the location of the possible counterpart
RX~J0546.8$-$6851 of the flaring source (see section \ref{sec:variability}).
\label{fig:30Dor}
}
\end{figure}

\begin{table}[!t]
\caption{Parameters of the point source model \psmod.
}
\label{tab:psmodel}
\begin{center}
\begin{tabular}{lrccrr}
\hline\hline
Name & \multicolumn{1}{c}{$\Delta$TS} & \ra\ & \dec\ & \multicolumn{1}{c}{$r95$} & \multicolumn{1}{c}{Flux} \\
\hline
S1 & 798.1 & $05^{\rm h}38.9^{\rm m}$ & $-69\deg17'$ & $4'$ & $11.8 \pm 1.8$ \\
S2 & 163.8 & $05^{\rm h}30.2^{\rm m}$ & $-67\deg16'$ & $12'$ & $7.6 \pm 2.1$ \\
S3 & 66.5 & $05^{\rm h}27.7^{\rm m}$ & $-68\deg58'$ & $10'$ & $4.0 \pm 1.7$ \\
S4 & 33.1 & $05^{\rm h}11.2^{\rm m}$ & $-67\deg15'$ & $25'$ & $3.9 \pm 2.3$ \\
S5 & 27.8 & $04^{\rm h}55.8^{\rm m}$ & $-69\deg21'$ & $11'$ & $2.6 \pm 1.4$ \\
\hline
\end{tabular}
\end{center}
Note to the table: Columns are
(1) source name,
(2) TS increase after including the respective point source in the model,
(3) Right Ascension of source,
(4) Declination of source,
(5) $95\%$ containment radius, and
(6) integrated $>$~100~MeV photon flux in units of $10^{-8}$ \funit\ assuming
a power law spectral slope with free index for each source.
Quoted uncertainties are statistical only.
\end{table}

Second, instead of using point sources we repeated the procedure with 2D Gaussian shaped
intensity profiles to build a geometrical model that is more appropriate for extended and 
diffuse emission structures.
We again stopped the successive addition of 2D Gaussian shaped sources once the TS
improvement after adding a further source dropped below 25.
This occurred after two 2D Gaussian shaped sources have been added to the model.
Table \ref{tab:gmodel} lists the maximum likelihood positions, Gaussian widths $\sigma$ and
fluxes for the two significant 2D Gaussian sources.
Similar to the point sources, the column $\Delta$TS quotes the TS improvements in each of
the successive steps.
The total TS amounts to 1122.6 for this model, corresponding to a detection significance 
of $32.7\sigma$ ($p=10$).
We refer to this model from now on as \gmod.

Adding a 3rd 2D Gaussian shaped source improved the TS by only 21.9, corresponding
to a detection significance of $3.5\sigma$ for this source ($p=5$).
We note that this 3rd 2D Gaussian source is located at
$(\ra, \dec) = (05^{\rm h}17.8^{\rm m}, -72\deg32')$ with a 95\% containment radius of
$18'$, and that it formally is consistent with a point source ($\sigma=10' \pm 13'$).
Using our standard source association procedure \citep{abdo09a} we found the blazar
CRATES~J051636$-$723707 as a possible counterpart of this source, suggesting
the presence of a background blazar within the boundaries of the LMC
(see also section \ref{sec:blazars}).

\begin{table}[!t]
\footnotesize
\caption{Parameters of the 2D Gaussian sources model \gmod.
Columns are identical to those in Table~\ref{tab:psmodel}, except for column 
6 which gives the width $\sigma$ for each 2D Gaussian source.}
\label{tab:gmodel}
\begin{center}
\begin{tabular}{lrccrrr}
\hline
\hline
\noalign{\smallskip}
Src. & \multicolumn{1}{c}{$\Delta$TS} & \ra\ & \dec\ & \multicolumn{1}{c}{$r95$} & \multicolumn{1}{c}{$\sigma$} & \multicolumn{1}{c}{Flux} \\
\noalign{\smallskip}
\hline
\noalign{\smallskip}
G1 & 1000.9 & $05^{\rm h}26.0^{\rm m}$ & $-68\deg16'$ & $20'$ & $73' \pm 5'$ & 
$19.6\pm2.2$ \\
G2 & 121.7 & $05^{\rm h}38.8^{\rm m}$ & $-69\deg18'$ & $7'$ & $12' \pm 4'$ & 
$8.5\pm2.2$ \\
\noalign{\smallskip}
\hline
\end{tabular}
\end{center}
\end{table}

\begin{figure*}[!th]
\centering
\includegraphics[width=7.6cm]{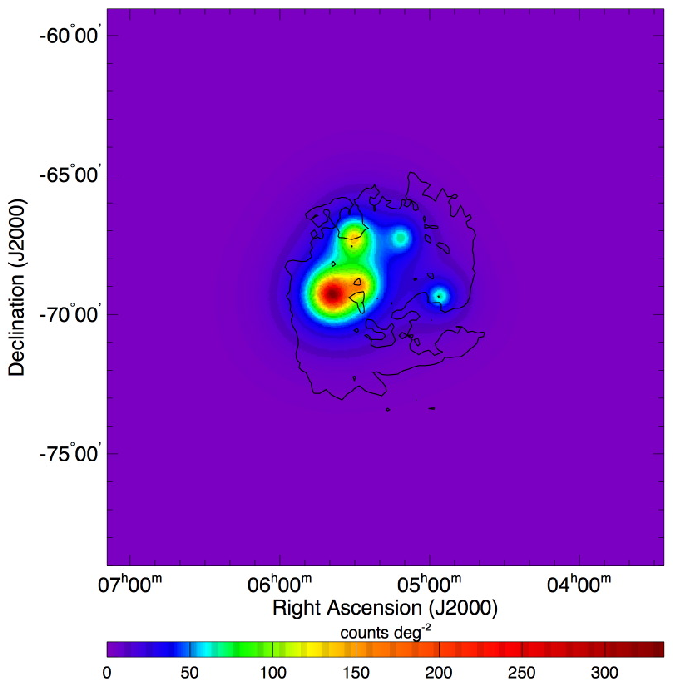}
\includegraphics[width=7.6cm]{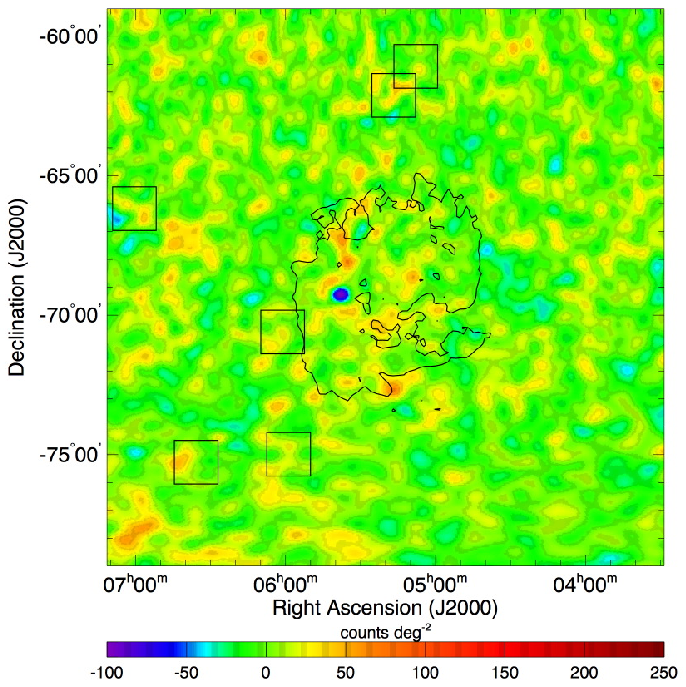}
\includegraphics[width=7.6cm]{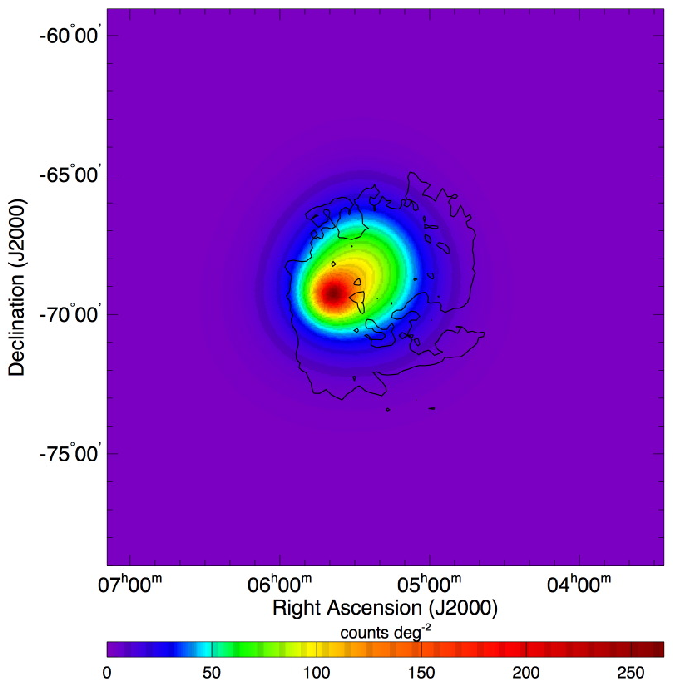}
\includegraphics[width=7.6cm]{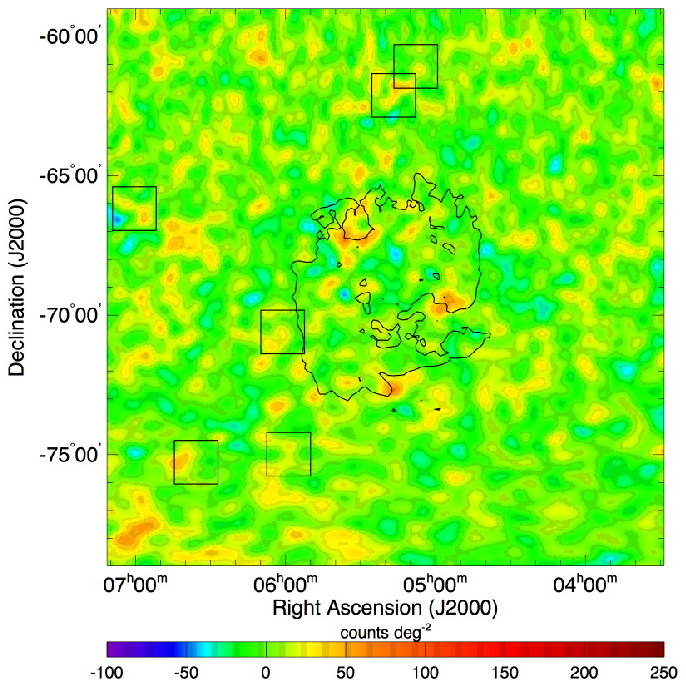}
\includegraphics[width=7.6cm]{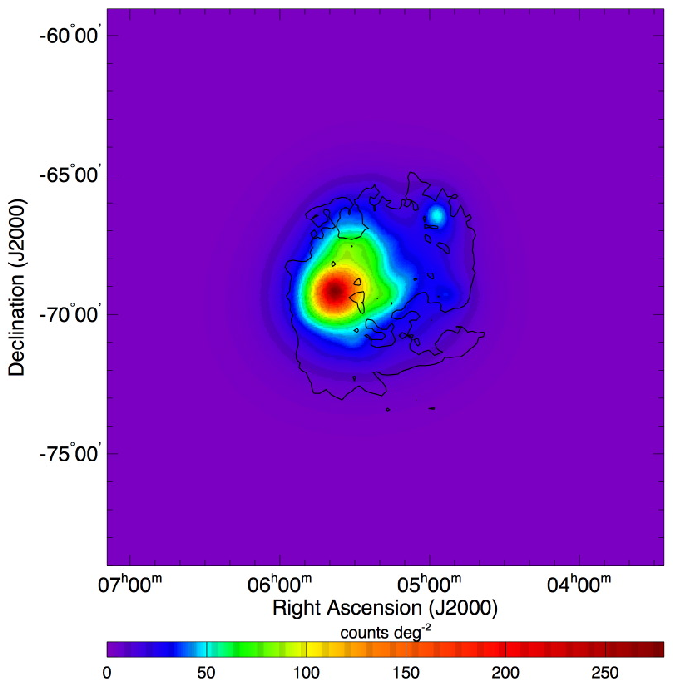}
\includegraphics[width=7.6cm]{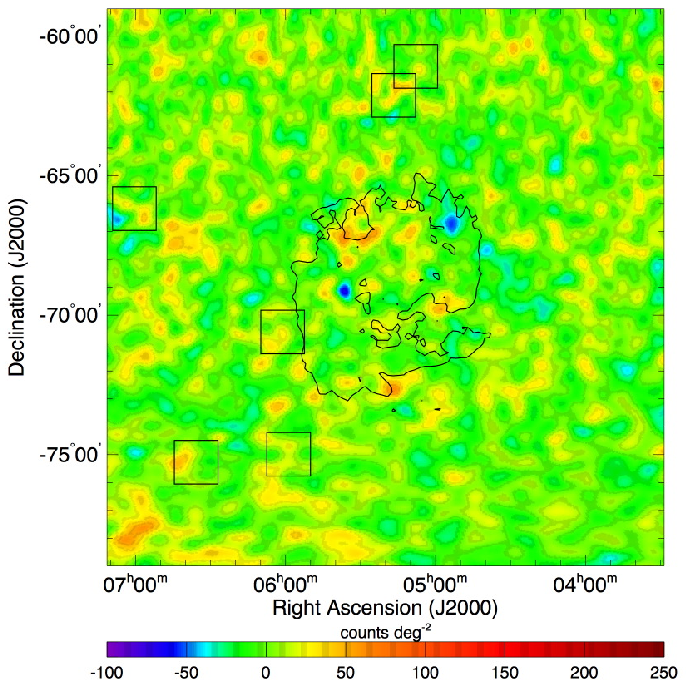}
\caption{
Gaussian kernel ($\sigma=0.2\deg$) smoothed model counts maps (left panels) and residual maps 
(right panels) of the region of interest (ROI).
The model counts maps have been obtained by convolving the model sky maps that were scaled 
to the fitted flux with the LAT point spread function.
The top row shows the \psmod\ model, the middle row shows the \gmod\ model and the
bottom row shows the \hii\ template.
The colour scaling and the dynamic range for the residual maps is identical to the right panel 
of Fig.~\ref{fig:image} to allow for comparison.
\label{fig:resmaps}
}
\end{figure*}

To illustrate how well the models fit the data we show in Fig.~\ref{fig:resmaps} the model
counts map and the residual counts that are left in the ROI after subtracting the \psmod\ or
\gmod\ model from the data.
The colour scale and dynamic range of the residual maps has been chosen identical to that of 
the right panel of Fig.~\ref{fig:image} to allow the comparison of the residuals before and 
after subtraction of the LMC model.
Apparently, only very few residual counts are left in the LMC area after subtraction of 
either of the two models from the data
(however a significant negative residual at the position of S1 indicates an overestimation
of the flux in the \psmod\ model in this area).
None of the peaks in the residuals is statistically significant ($\Delta$TS~$<25$)
as a point source (for the \psmod\ model) or a 2D Gaussian shaped extended source 
(for the \gmod\ model).

The \psmod\ and \gmod\ models have in common that they both comprise a strong 
source located at
$(\ra, \dec) \approx (05^{\rm h}39^{\rm m}, -69\deg17')$ in the 30~Dor region
(S1 for \psmod\ and G2 for \gmod).
We indicate the location uncertainties for these sources as circles in Fig.~\ref{fig:30Dor}.
We also show the locations of the powerful star cluster R~136 and of the two Crab-like
pulsars \psra\ and \psrb.
From this figure it becomes clear that S1 in the \psmod\ model does not coincide with any 
of these objects.
We confirm this observation by moving S1 to the locations of either \psra, \psrb\ or 
R~136, which leads to a significant TS reduction of
21.0 ($4.2\sigma$) for the position of \psra,
28.3 ($4.9\sigma$) for the position of \psrb, and
31.1 ($5.2\sigma$) for the position of R~136.
Thus, none of the 3 sources can explain the observed gamma-ray emission from
30~Dor alone.
We come to the same conclusion when we replace G2 in the \gmod\ model by a point source
at any of the 3 locations.

To search for evidence of point source emission from \psra, \psrb\ or R~136 on top
of the diffuse emission, we added point sources at the respective positions to the \gmod\ model.
Adding any of the sources led to only marginal TS improvements of
0.5 ($0.3\sigma$) for \psra,
0 ($0\sigma$) for \psrb, and
0.1 ($0.1\sigma$) for R~136.
Adding all three sources together marginally improved the TS by 0.5 ($0.003\sigma$).
Thus, we found no evidence for point source emission from any of the 3 sources on top
of the diffuse emission that is modelled by G2.
When we replaced G2 by a combination of 3 point sources located at the positions
of \psra, \psrb\ and R~136 we obtained a TS of 1113.4, which still is smaller by 
$\Delta$TS~$=9.2$ 
than the TS obtained for the \gmod\ model.
Obviously, the gamma-ray emission from 30~Dor is better described by a single
2D Gaussian shaped extended source than by a combination of point sources
situated at the locations of \psra, \psrb\ and R~136.

\subsubsection{Gas templates}
\label{sec:templatefitting}

To test specific hypotheses about the spatial distribution of the gamma-ray intensity 
we compared our data to spatial templates that trace the interstellar matter distribution
in the LMC.
The reasoning behind this comparison is that gamma-rays are expected to primarily
arise from interactions between cosmic rays and the interstellar medium 
\citep{ozel88}, 
and we thus may obtain a measurement of the cosmic-ray density variations from the
gamma-ray to gas ratio.

Most of the gas in the LMC is found in form of neutral atomic hydrogen and helium
\citep{staveleysmith03} while $5\%-10\%$ in mass is in form of molecular 
clouds \citep{fukui08} and about $1\%$ in mass is in form of ionized hydrogen 
\citep{kennicutt95}.
For the atomic hydrogen component (\hi) we used the map of \citet{kim05} that is based
on a combination of aperture synthesis and multibeam data from ATCA and Parkes 
observations.
We converted this map into hydrogen column densities assuming optically thin emission
or by applying an optical depth correction under the assumption of uniform spin 
temperature of either 40~K, 60~K or 125~K \citep{marxzimmer00}.
\citet{bernard08} suggested the presence of an additional dark gas component in the LMC
which correlates well with the distribution of the observed \hi.
Thus, if this dark gas component indeed exists it should be morphologically well 
described by the \hi\ templates.
For the molecular hydrogen component (\hmol) we used the CO map of the LMC
obtained with the NANTEN telescope \citep{fukui08} which we converted into molecular
hydrogen column densities using a constant CO intensity to H$_2$ column density 
conversion factor 
$X_{\rm CO} = 7\times 10^{20}$ H$_2$~cm$^{-2}$ / (K~km~s$^{-1}$)
\citep{fukui08}.
For the ionized hydrogen component we used the full-sky composite \halpha\ map
of \citet{finkbeiner03} which we converted into ionized hydrogen column densities
using
$N(\hii) = 1.37 \times 10^{18} I_{{\rm H}\alpha}$
which assumes a gas temperature of 8000~K and an electron density
of $n_e=1$~cm$^{-3}$ \citep[cf.~Eq.~6 of][]{bernard08}.
In addition to the individual templates for the interstellar gas phases we also fitted
a template of the total gas column density $N({\rm H})$ to the data that we derived by 
adding together the column densities of \hi, \hmol, and \hii.
Furthermore, we also fitted the linear combination $N(\hi)+N(\hmol)+N(\hii)$ to the data
with independent scaling factors and power law indices for each component.

The TS values and detection significances that are obtained by maximum likelihood 
fitting are summarized in Table~\ref{tab:models}, where we also give the corresponding
results for the geometrical models \psmod\ and \gmod\ for reference.
Fitting the optically thin \hi\ and \hmol\ maps resulted in TS values that are considerably
worse than those obtained for the geometrical models.
Applying the optical depth correction to the \hi\ data slightly improved the
fit with a maximum TS of $\sim835$ that is reached for spin temperatures of
40~K - 60~K, yet still, this TS is considerably worse than that of the geometrical models.
Apparently, the \hi\ and \hmol\ maps provide rather poor fits to the data, indicating 
that the distribution of gamma rays does not follow the distribution of neutral hydrogen 
in the LMC.

\begin{table}[!t]
\footnotesize
\caption{Comparison of maximum likelihood model fitting results (see text for a
description of the models).}
\label{tab:models}
\begin{center}
\begin{tabular}{lrcr}
\hline\hline
Model & \multicolumn{1}{c}{TS} & Significance & \multicolumn{1}{c}{$p$} \\
\hline
\psmod\ & 1089.3 & 31.5 & 20 \\
\gmod\ & 1122.6 & 32.7 & 10 \\
\hline
$N(\hi)$ (optically thin) & 771.8 & 27.6 & 2 \\
$N(\hi)$ (D+L components) & 792.9 & 27.8 & 4 \\
$N(\hi)$ ($T_s=125$~K) & 794.9 & 28.1 & 2 \\
$N(\hi)$ ($T_s=60$~K) & 834.6 & 28.8 & 2 \\
$N(\hi)$ ($T_s=40$~K) & 835.3 & 28.8 & 2 \\
$N(\hmol)$ & 824.3 & 28.6 & 2 \\
$N(\hii)$ & 1110.1 & 33.2 & 2 \\
$N({\rm H})$ & 806.9 & 28.3 & 2 \\  
$N(\hi)+N(\hmol)+N(\hii)$ & 1110.7 & 32.9 & 6 \\
\hline
\end{tabular}
\end{center}
Note to the table:
Columns are
(1) the name of the model,
(2) the TS of the fit,
(3) the detection significance, and
(4) the number of free parameters.
\end{table}

\citet{luks92} proposed that the neutral hydrogen in the LMC is confined into two separate
structural features (dubbed 'D' and 'L' components), and we tested whether gamma-ray emission 
is possibly only correlated to one of these \hi\ components.
We did this by separating the \hi\ data of \citet{kim05} into two maps covering the heliocentric
radial velocity intervals 
$190<V_{\rm rad}<270$ km~s$^{-1}$ (L) and
$270<V_{\rm rad}<386$ km~s$^{-1}$ (D) that roughly separate the two components
\citep[see Fig.~4 of][]{luks92}.
Fitting both maps simultaneously with independent scaling factors and power law indices
to our data only marginally improved the fit upon
the $N(\hi)$ gas map ($\Delta$TS~$=21.1$).
We therefore conclude that the gamma-ray emission does also not follow the distribution
of either the D or the L component identified by \citet{luks92}.

The \hii\ map, on the other hand, gave a TS that is very close to that of the \psmod\ and 
\gmod\ models.
Fitting the total hydrogen column densities, i.e.~model $N({\rm H})$, gave a rather low
TS, close to that obtained for the \hi\ and \hmol\ templates alone.
This result is explained by the preponderance of atomic hydrogen in the LMC which 
means that the maps of total hydrogen and \hi\ are very similar.
Fitting a linear combination of the three gas phases, i.e.~model $N(\hi)+N(\hmol)+N(\hii)$,
gave basically the same TS value that is obtained for the \hii\ template alone.
In this case, \hii\ is in fact the only significant model component in the fit while
the maximum likelihood fluxes for the other components are negligible.

The \hii\ map thus provided the best fit among all of the gas maps to the LAT data.
Figure~\ref{fig:resmaps} shows the model counts map for the \hii\ template and the
corresponding residual counts map, which apparently is similar to the residual maps that 
we obtain for the geometrical models.
A search for point-like or extended emission on top of the \hii\ template did 
not reveal any significant additional signal ($\Delta$TS~$<25$).
The \hii\ map is characterised by a strong emission peak near 30~Dor which is attributed
to the intense ionizing radiation of the massive stars in this highly active region.
To test whether the fits are mainly driven by this peak of ionizing flux, we repeated the template
map analysis by adding the source G2 which describes the emission from 30~Dor in the
\gmod\ model as an additional component to our background model.
The flux and spectral index of this component were kept as free parameters.
At the same time, we modified the gas maps by setting all pixels within a circular region of 
$0.5\deg$ around $(\ra,\dec)=(05^{\rm h}38^{\rm m}42^{\rm s}, -69\deg06'03")$
to zero which removed any peak associated to 30~Dor from the templates (in particular,
the bright peak of ionizing flux in 30~Dor is now removed).
In that way, we considerably reduced the impact of the gamma-ray emission from 30~Dor
on the fit of the gas templates.

\begin{table}[!t]
\footnotesize
\caption{Comparison of maximum likelihood model fitting results after inclusion of
the source G2 in the background model.
Columns are identical to those in Table~\ref{tab:models}.}
\label{tab:g2models}
\begin{center}
\begin{tabular}{lccc}
\hline\hline
Model & TS$^\prime$ & Significance & $p$ \\
\hline
$N'(\hi)$ (optically thin) & 173.1 & 12.9 & 2 \\
$N'(\hi)$ (D+L components) & 194.1 & 13.7 & 2 \\
$N'(\hi)$ ($T_s=125$~K) & 175.4 & 13.0 & 2 \\
$N'(\hi)$ ($T_s=60$~K) & 180.0 & 13.2 & 2 \\
$N'(\hi)$ ($T_s=40$~K) & 178.4 & 13.1 & 2 \\
$N'(\hmol)$ & 197.0 & 13.9 & 2 \\
$N'(\hii)$ & 284.9 & 16.7 & 2 \\
$N'({\rm H})$ & 183.7 & 13.3 & 2 \\
$N'(\hi)+N'(\hmol)+N'(\hii)$ & 284.9 & 16.1 & 6 \\
\hline
\end{tabular}
\end{center}
\end{table}

We summarise the results of this analysis in Table \ref{tab:g2models}, where the prime
in the model names indicates that the 30~Dor emission has been removed from the
gas templates.
The TS values are now considerably reduced since $\mathcal{L}_0$ now includes the
G2 source that accounts for gamma-ray emission from 30~Dor.
Consequently, TS now measures the significance of gamma-ray emission that exists 
in addition to that seen towards 30~Dor in the LMC.
To recognise this change in the definition we label the corresponding column in Table
\ref{tab:g2models} by TS$^\prime$.\footnote{
 Fitting the G2 source as the only component to the data resulted in TS$=826.5$.
 We can thus formally convert TS$^\prime$ into TS using TS$=$TS$^\prime+826.5$.}
Globally, however, we found the same trend as before:
the \hii\ template provided the best fit to the data while the \hi\ and \hmol\ templates
resulted in significantly smaller TS$^\prime$ values.
We thus conclude that the morphology of the gamma-ray emission from the LMC
is not well explained by the distribution of neutral gas in that galaxy, irrespectively of 
whether the emission from 30~Dor is taken into account.
A template based on the distribution of ionized hydrogen as measured by the \halpha\
emission describes the spatial distribution of the gamma-ray emission in the LMC
considerably better than the distribution of neutral gas.

\subsection{Variability}
\label{sec:variability}

As we noted earlier (cf.~section \ref{sec:backgroundmodel}), our data indicated possible
flaring activity within the LMC boundaries during month 4 (MJD 54777.8 - 54808.2)
of the observation interval, and we excluded the data of this month from our analysis.
Now with reliable spatial templates for the LMC gamma-ray emission at hand
(the \gmod\ model and the \hii\ gas map), we reconsidered the 
time variability of the gamma-ray emission in the LMC area.

\begin{figure}[!t]
\centering
\includegraphics[width=9.1cm]{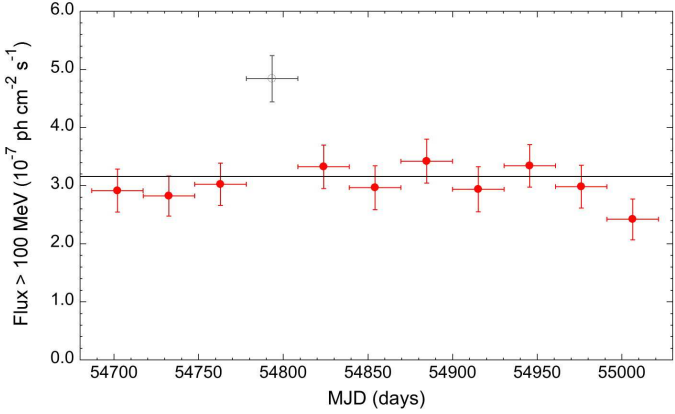}
\caption{
Monthly lightcurve for gamma-ray emission from the LMC obtained using the \hii\ 
gas map.
The 4th month of data that was excluded from our analysis is indicated by an open
symbol and grey colour.
The solid line indicates the average flux over the full 11 months of data that is obtained
by assuming a power law spectral shape for the emission.
\label{fig:lightcurve}
}
\end{figure}

First, we considered the time-variability of the integrated gamma-ray emission from the 
LMC.
For this purpose we fitted the \gmod\ model and the \hii\ gas map to the data on a monthly, 
2 weeks and weekly basis using a power law to describe the source spectrum.
We fixed the power law spectral index to the average value that we obtain from fitting
the \gmod\ model or the \hii\ gas map to all data excluding month 4.
Figure \ref{fig:lightcurve} shows the monthly lightcurve that we obtain using the \hii\ 
gas map.
Fitting the lightcurve with a constant flux level (solid line) resulted in $\chi^2=25.4$ for 10 
degrees of freedom, which corresponds to a probability of $0.2\%$ for the flux being
constant.
Obviously, the flux measured during month 4 (MJD 54777.8 - 54808.2) from the LMC
is significantly larger than the flux measured for the remaining period.
Similar results were obtained when we perform the variability analysis on time scales of
2 weeks ($\chi^2=38.9$ for 23 degrees of freedom, $1.0\%$ probability for constant 
flux) and
1 week ($\chi^2=69.9$ for 47 degrees of freedom, $0.8\%$ probability for constant 
flux).
Using the \gmod\ models instead of the \hii\ gas map gave comparable results.

As next step we tried to localise any flaring source (and in particular the one that occurred
during month 4) by fitting a point source with free position, flux and spectral power law 
index on top of the \gmod\ model and the \hii\ gas map to each of the 11 monthly datasets.
We fixed the flux and spectral index of the \gmod\ model and \hii\ gas map to the values
that we obtained from the fit to all data excluding month 4.
This analysis revealed that month 4 is the only period where a point source is detected
significantly (TS~$>25$) on top of the LMC models within the galaxy's boundaries.
Using the \hii\ map as template for the LMC, we localised the source at
$(\ra,\dec)=(05^{\rm h}46.4^{\rm m},-69\deg01')$ with a 95\% containment radius of $29'$
and a detection significance of $4.6\sigma$ (TS~$=30.1$),
while with the \gmod\ model we found
$(\ra,\dec)=(05^{\rm h}46.2^{\rm m},-69\deg01')$ with a 95\% containment radius of $36'$,
and a detection significance of $4.5\sigma$ (TS~$=29.3$).
Obviously, the choice of the LMC model had a negligible impact on the maximum
likelihood location of the source.

We show the larger localisation error circle obtained with the \gmod\ model in 
Fig.~\ref{fig:30Dor}, which illustrates that the flaring source is located near 30~Dor and 
that it is close to the maximum gamma-ray intensity observed from the LMC.
Independent on the LMC model, we found an integrated $>100$~MeV photon flux of
$(21 \pm 8~({\rm stat}) \pm 4~({\rm sys})) \times 10^{-8}$ \funit\
for the flaring source during month 4 and obtained a maximum likelihood power law
spectral index of
$\Gamma = 2.8 \pm 0.3~({\rm stat}) \pm 0.1~({\rm sys})$.
The spectral index is rather steep and compatible with the softest indices that are typically
found for flat-spectrum radio quasars (FSRQs) \citep{abdo09b}.
FSRQs tend to show also the largest variabilities among all source classes detected by
LAT \citep{abdo09a,abdo09b} which makes them good candidates for flaring sources.
Using the source association procedure described by \citet{abdo09a}, we searched for
plausible counterparts of the flaring source.
We did not find a plausible counterpart in any of our standard blazar or AGN
catalogues.
The only object that fulfilled our association criterion was RX~J0546.8$-$6851 which is listed in
the Magellanic Cloud high-mass X-ray binary catalogue of \citet{liu05}.
However, the nature of this source that has been discovered by the {\em Einstein Observatory} 
\citep{wang91} is not yet established.
So far, no companion star has been identified for RX~J0546.8$-$6851, and following
\citet{kahabka01} the source could also be an active galactic nucleus.
It is thus conceivable that the flaring source detected by LAT during month 4 is not physically
associated with the LMC, but is rather a yet unidentified background blazar that by chance is 
located near the line of sight toward 30~Dor.

\subsection{Blazars in the field of the LMC}
\label{sec:blazars}

As we noted earlier (cf.~section \ref{sec:backgroundmodel}), we expect to find 1-2 gamma-ray
blazars within the boundaries of the LMC in our dataset.
Identification of these blazars would be possible if they were flaring during the observation
period (see the previous section), yet a substantial fraction of the blazars seen by the LAT
do not show any significant time variability \citep{abdo09b}.
Blazars are thus difficult to recognize if they spatially concur with gamma-ray emission from 
the LMC.
Nevertheless, we can try to spot blazar candidates by searching for point source emission 
that spatially coincides with the locations of blazars.

\begin{figure}[!t]
\centering
\includegraphics[width=9.1cm]{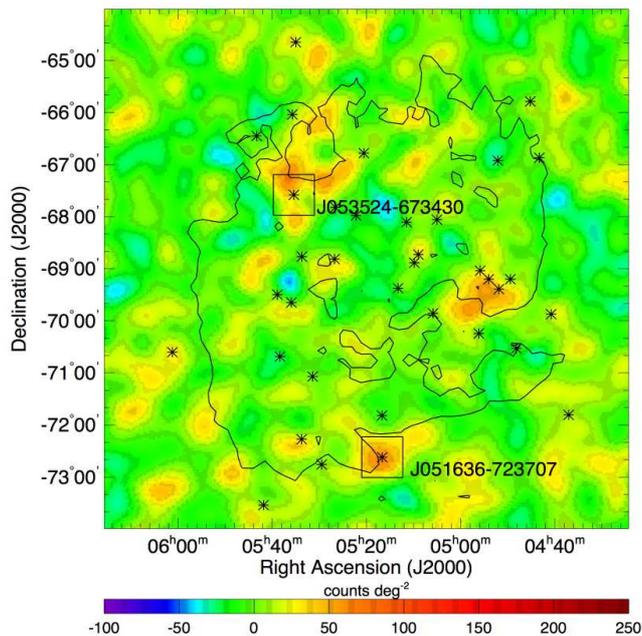}
\caption{
Gaussian kernel ($\sigma=0.2\deg$) smoothed residual map for the \gmod\ model over
which we overplot the CRATES blazars in the field of the LMC as stars.
The boxes mark those blazars for which TS~$>6$.
\label{fig:blazars}
}
\end{figure}

For this purpose we searched for counterparts for any of the point sources in the \psmod\ 
model. 
Using our standard source association procedure \citep{abdo09a} we did not find any
blazar candidate coincident with any of the sources of the \psmod\ model.

\begin{figure*}[!th]
\centering
\includegraphics[width=9.1cm]{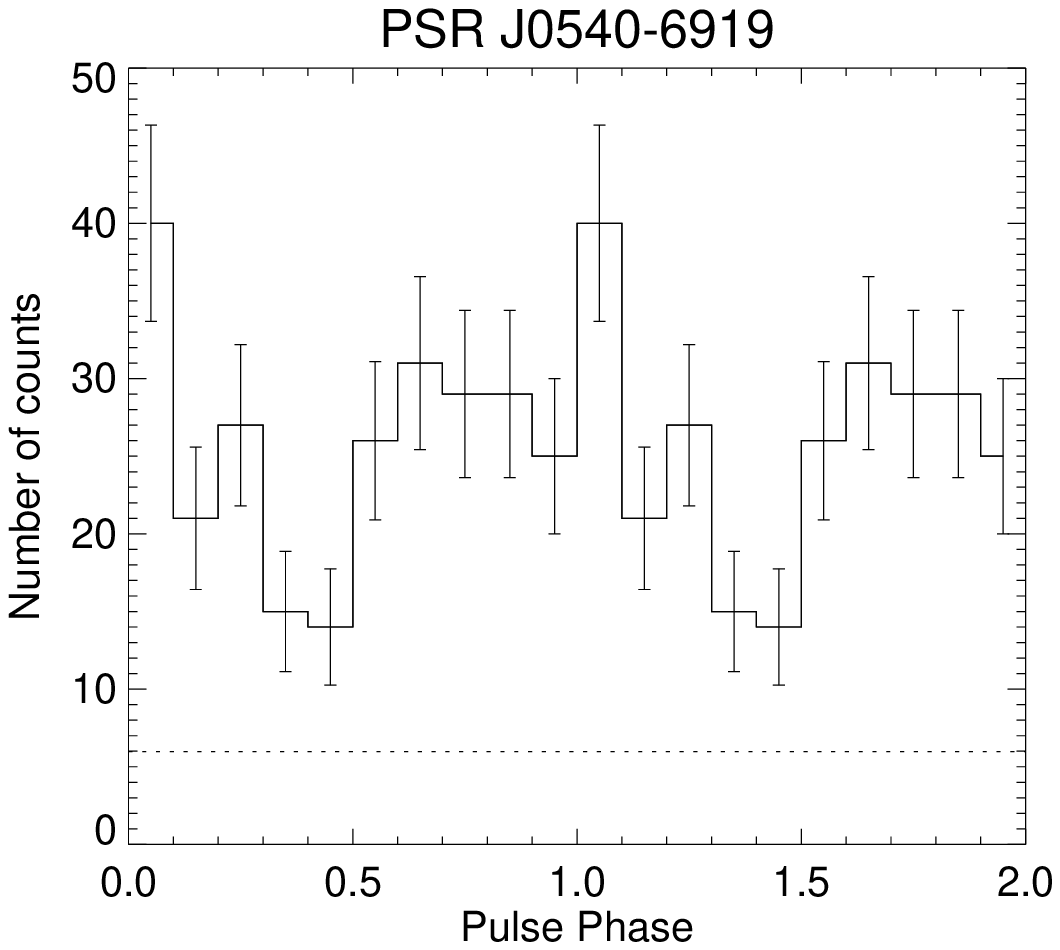}
\includegraphics[width=9.1cm]{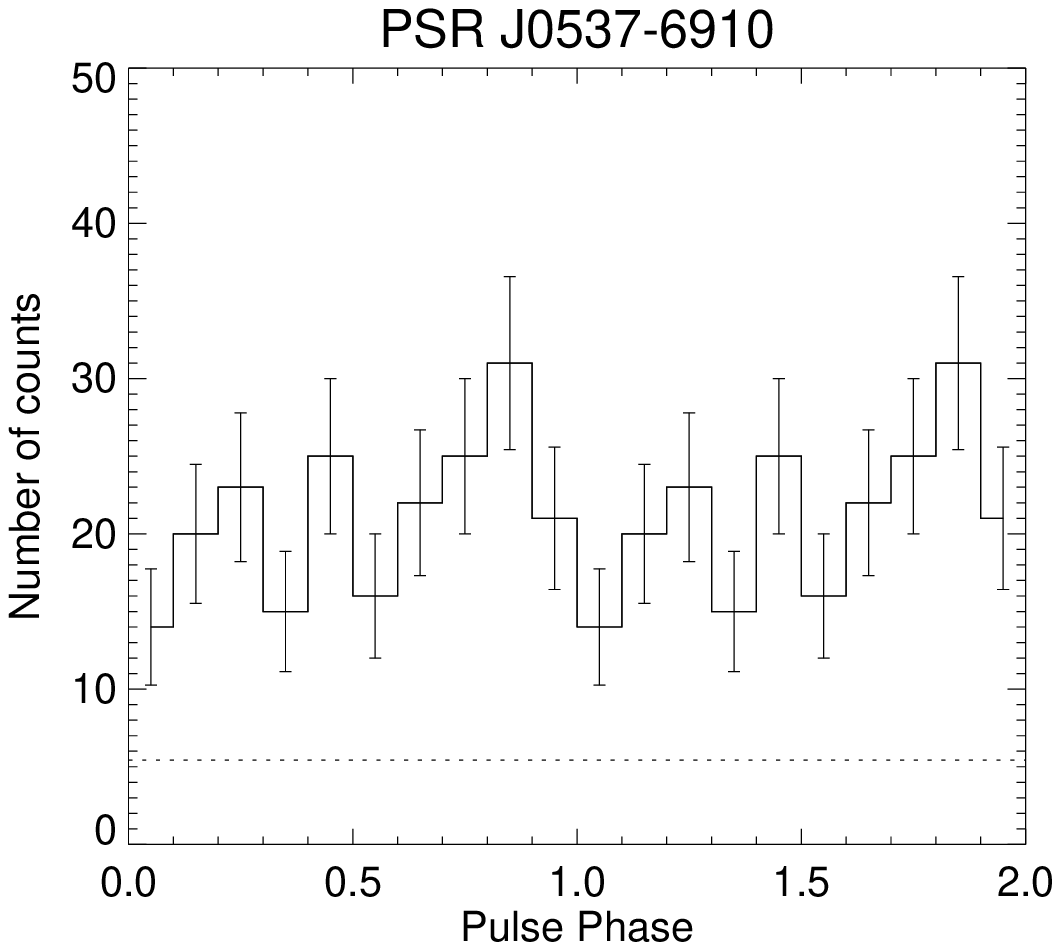}
\caption{
Lightcurves obtained for \psra\ (left) and \psrb\ (right) for the
energy range 200 MeV - 20 GeV using an event selection radius of $0.5\deg$ around 
the nominal pulsar positions.
The data are binned into 10 phase bins.
The dotted lines indicate the sum of the expected contributions from
extragalactic and Galactic diffuse emissions
and residual instrumental background.
\label{fig:pulsar}
}
\end{figure*}

As next step, we simultaneously fitted point sources at the positions of all CRATES blazars
\citep{healey07} that are found within a radius of $5\deg$ around the centre of the LMC,
which we assumed here to be
$(\ra, \dec)=(05^{\rm h}17^{\rm m}36^{\rm s}, -69\deg01'48")$,
in addition to the \gmod\ model and the \hii\ template to the data.
The fluxes and the spectral power law indices of all sources were left as free parameters
of the fit.
We found a total of 36 CRATES sources within the selection region of which one
(CRATES J060106$-$703606) was already part of our background model
(see Table~\ref{tab:crates}).
This left us with 35 CRATES blazars in the field.
As expected from our previous searches for significant point sources above the \gmod\
model and the \hii\ template (cf.~sections \ref{sec:geomodels} and \ref{sec:templatefitting}), 
none of these blazars was significantly detected (TS~$>25$).
However, two sources had a formal detection significance above $2\sigma$:\footnote{
which corresponds to TS~$\ga6$ for $p=2$.}
CRATES J051636$-$723707 (TS~$=16.7$), and
CRATES J053524$-$673430 (TS~$=12.4$).
The locations of these blazars are overplotted together with those of all other CRATES sources
in the LMC field over the residual map for the \gmod\ model in Fig.~\ref{fig:blazars}.

CRATES J051636$-$723707 appeared to be point like and had already been associated in
section \ref{sec:geomodels} as plausible counterpart of the 3rd 2D Gaussian in the \gmod\
model analysis.
Figure \ref{fig:blazars} clearly shows that this source is associated with a peak in the observed
counts maps.
CRATES J053524$-$673430, in contrast, is not associated with any emission peak but is rather
situated in the centre of an extended emission feature.
Fitting a 2D Gaussian shaped emission profile at the location of CRATES J053524$-$673430
suggested indeed that the emission is extended ($\sigma=26' \pm 10'$), making the blazar an
unlikely counterpart of the gamma-ray emission in that area.
CRATES J051636$-$723707 seems to be thus the only plausible blazar candidate in the field
of the LMC.

We note, however, that although many of the LAT high-latitude sources have counterparts
in the CRATES catalogue, there is still a non-negligible population of unassociated high-latitude
sources in the LAT source list that may consist of yet unknown (or unidentified) blazars.
Some of the residuals seen in Fig.~\ref{fig:blazars} could indeed arise from such blazars, yet 
their emission does not appear to significantly contribute to the overall emission from the
LMC region.

\subsection{Pulsar lightcurves}
\label{sec:pulsarlc}

As we noted earlier (cf. section \ref{sec:spatial} and Fig.~\ref{fig:30Dor}), 
the peak of gamma-ray emission 
that we see toward 30~Dor is spatially close to the locations of the powerful young 
pulsars \psra\ and \psrb, which both are potential sources of high-energy gamma rays.
By searching for their pulsations in our data we can assess if the pulsars indeed
contribute to the observed emission.
We used for this search ephemerides that have been obtained from concurrent {\em RXTE}
observations (Marshall, private communication) and that cover the time intervals
MJD 54651-55015 for \psra\ and
MJD 54710-54714, 54751-54771, and 54885-54897 for \psrb.
Only data within these time intervals have been used for the analysis.
Gamma-ray photon arrival times were referred to the solar-system barycentre and pulse 
phases were assigned using the standard pulsar timing software TEMPO2 
\citep{hobbs06}.

Figure \ref{fig:pulsar} shows the lightcurves that we obtained by selecting photons within
a radius of $0.5\deg$ around the nominal pulsar locations for the energy range
200 MeV - 20 GeV.
For \psra\ our data may indicate the presence of pulsations, but the H-test statistic amounts
to only 10.1 which corresponds to a detection significance of $2.4\sigma$
for the pulsations.
For \psrb\ the lightcurve is featureless and an H-test gives
3.2 corresponding to a detection significance of $1.1\sigma$.
Thus, neither pulsar is detected significantly in our data which is in line with our
previous finding (cf.~section  \ref{sec:geomodels}) that the gamma-ray emission from
30~Dor cannot be satisfactorily described by point source emission from the pulsars.

\subsection{Spectrum}

\begin{figure}[!t]
\centering
\includegraphics[width=9.1cm]{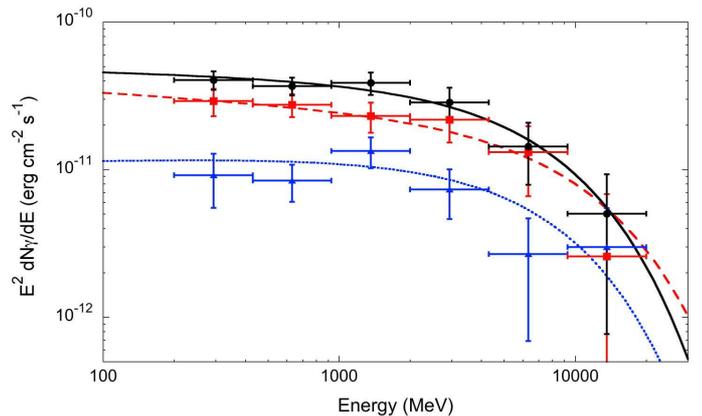}
\caption{
Spectrum of the LMC emission obtained by fitting spatial model maps in 6 logarithmically 
spaced energy bins covering the energy range 200 MeV - 20 GeV.
Black dots show the total spectrum obtained by fitting the \hii\ gas map,
red squares and blue triangles show the spectra for the LMC disk and for 30~Dor,
respectively, obtained using the \gmod\ model.
Error bars include statistical and systematic uncertainties.
The lines show the results obtained by fitting exponentially cut off power law spectral
models to the data using maximum likelihood fits.
\label{fig:spectrum}
}
\end{figure}

So far, all analysis has been done by assuming that the spectrum of the LMC emission components
is well described by power laws.
To determine the spectrum of the gamma-ray emission 
from the LMC independently of any assumption on the spectral shape, we fitted our data in  
6 logarithmically spaced energy bins covering the energy range 200 MeV - 20 GeV.
We obtained the total spectrum of the LMC by fitting the \hii\ template to the data.
We also obtained separate spectra for the LMC disk and for 30~Dor by fitting 
the \gmod\ model to the data.
Here, G1 is taken to represent the LMC disk emission while G2 is taken to represent the 
emission from 30~Dor.
The results are shown in Fig.~\ref{fig:spectrum}.
The spectra are relatively flat below a few GeV ($\Gamma \sim 2$) and show evidence
for a break or cut off above that energy.

\begin{table*}[!t]
\footnotesize
\caption{
Maximum likelihood spectral fit results.
\label{tab:spectralfits}
}
\begin{center}
\begin{tabular}{lrrr}
\hline\hline
Parameter & \multicolumn{1}{c}{Total} & \multicolumn{1}{c}{Disk} & \multicolumn{1}{c}{30~Doradus} \\
\hline
Photon flux ($>$100 MeV) & $26.3\pm2.0\pm4.2$ & $18.4\pm2.3\pm3.0$ & $6.9\pm1.7\pm0.7$ \\
Energy flux ($>$100 MeV) & $16.0\pm0.6\pm1.8$ & $11.1\pm0.7\pm1.5$ & $4.6\pm0.5\pm0.5$ \\
Photon flux ($>$200 MeV) & $12.2\pm0.4\pm0.7$ & $8.3\pm0.6\pm0.7$ & $3.3\pm0.4\pm0.3$ \\
Energy flux ($>$200 MeV) & $12.9\pm0.4\pm1.8$ & $8.9\pm0.5\pm1.6$ &  $3.8\pm0.3\pm0.5$ \\
$\Gamma$ & $2.04\pm0.09\pm0.07$ & $2.10\pm0.13\pm0.10$ & $1.96\pm0.25\pm0.02$ \\
\ecut\ (GeV) & $7.1^{+3.5+6.6}_{-2.1-2.4}$ & $10.3^{+12.4+\infty}_{-4.4-10.3}$ & $6.9^{+10.5+1.3}_{-3.5-1.0}$ \\
TS with respect to power law & 11.3 & 3.0 & 3.6 \\
Significance of cut off ($\sigma$) & 3.4 & 1.7 & 1.9 \\
\hline
\end{tabular}
\end{center}
Note to the table:
Photon fluxes are given in units of $10^{-8}$ \funit,
energy fluxes are given in units of $10^{-11}$ \eunit.
Errors are statistical and systematic.
\end{table*}

To determine the integrated gamma-ray flux of the spectra we fitted exponentially cut off power 
law spectral models of the form
$N(E) = k\,(E/E_0)^{-\Gamma} \exp(-E/\ecut)$
to the data.
We made these fits by means of a binned maximum likelihood analysis over the energy range 
200 MeV - 20 GeV.
This analysis is more reliable than fitting the spectra of Fig.~\ref{fig:spectrum} directly since it
accounts for the Poissonian statistic of the data.
The resulting spectral shapes are shown as lines in Fig.~\ref{fig:spectrum} and spectral
fit parameters are given in Table \ref{tab:spectralfits}.
Integrated fluxes are derived for energies above 100~MeV and 200~MeV by integrating the
spectral model numerically from the lower boundary up to 500~GeV.
The $>100$~MeV fluxes are mainly given for comparison with the former measurements done
by EGRET.
Since we restricted our analysis to photons $>200$~MeV the extrapolation down to 100~MeV
introduces an additional systematic uncertainty that for the photon flux can reach up to $30\%$
(cf.~Table \ref{tab:spectralfits}).
We also determined the significance of the spectral cut off by computing the TS difference
between fitting a simple power law and fitting an exponentially cut off power law for each
component.

\section{Discussion}
\label{sec:discussion}

\subsection{Comparison to EGRET}

\citet{sreekumar92} reported the first detection of the LMC in high-energy gamma rays
based on 4 weeks of data collected with the EGRET telescope aboard {\em CGRO}.
They obtained an integrated $>100$~MeV flux of
$(1.9 \pm 0.4) \times 10^{-7}$ \funit\
for the LMC from their data which is in reasonable agreement with our value of
$(2.6 \pm 0.2~{\rm(stat)} \pm 0.4~{\rm (sys)}) \times 10^{-7}$ \funit\
that we obtained from spectral model fitting.

The 3rd EGRET catalogue \citep{hartman99} gives positions, fluxes and power law spectral 
indices for point sources detected during 4 years of EGRET observations
and locates the LMC emission at
$(\ra, \dec)=(05^{\rm h}33^{\rm m}, -69\deg16')$ with a 95\% containment radius
of $32'$, close to our maximum likelihood location for S1 in the \psmod\ model or
G2 in the \gmod\ model, and consistent with the location of 30~Dor.
The $>100$~MeV photon flux in the 3rd EGRET catalogue amounts to 
$(1.4 \pm 0.2) \times 10^{-7}$ \funit,
which is significantly lower than our value.
The assumption of a single point source for the EGRET catalogue analysis may
have led to an underestimation of the flux from the extended LMC galaxy.
On the other hand, our $>100$~MeV flux is based on power-law extrapolation from our
measurement $>200$~MeV which could lead to overestimation at lower
energies.
The 3rd EGRET catalogue quotes a power law spectral index of $\Gamma = 2.2 \pm 0.2$
for the LMC which is consistent with our value.
In particular, fitting the total LMC emission with a simple power law instead of an exponentially 
cut off power law we obtain 
$\Gamma = 2.26 \pm 0.04~{\rm(stat)} \pm 0.10~{\rm (sys)}$, 
consistent with the value obtained from EGRET data.

Thus, globally, our results are in line with those found earlier by EGRET, yet we benefit from
a much better photon statistics\footnote{
  The total number of counts we attribute to the LMC amounts to $\sim1550$ photons while
  the 3rd EGRET catalogue quotes a total of 192 counts obtained during 4 years of observations.},
an improved angular resolution and spectral coverage up to at least $\sim10$~GeV.

\subsection{On the origin of gamma rays from 30~Doradus}
\label{sec:30dor}

Although it was already obvious from EGRET observations that the LMC was an
extended source, only the {\em Fermi}/LAT data allow now for the first time to clearly resolve the 
LMC and to localise the emission maximum near the 30~Dor massive star forming region.
Even if we cannot establish either of the pulsars \psra\ or \psrb\ as the source of the
gamma-ray emission toward 30~Dor (cf. sections \ref{sec:geomodels} and \ref{sec:pulsarlc}), 
the indication of possible pulsations from \psra\ is a challenging possibility.
If further observations of the LMC by {\em Fermi}/LAT confirm the detection of pulsations, this 
would be the first time that a pulsar outside our own Galaxy were detected
in gamma rays.
We are aware that the H.E.S.S. collaboration has announced\footnote{
The detection has been announced at the 31$^{\rm st}$ International Cosmic-Ray
Conference in {\L}\'{o}d\'{z} 2009, however no details about source flux and spectrum are yet
publicly available.
}
the detection of the pulsar wind nebula of \psrb\ at TeV energies, yet so far we do not find
evidence for gamma-ray emission from that source in our data.

As the gamma-ray emission from 30~Dor is not dominated by pulsars, cosmic-ray interactions 
with the ambient interstellar medium and radiation fields are plausibly the origins of the
observed emission.
In this case, the small size of the emission region puts stringent constraints on the 
diffusion length of GeV cosmic-ray protons.
For source G2 we find an extent of $\sigma=12'\pm4'$ and a $95\%$ confidence
level upper limit of $\sigma=19'$, which at the distance of the LMC corresponds to
linear dimensions of 170 pc and 270 pc, respectively.
This is extremely small compared to expectations.
Gamma-ray observations of our own Galaxy suggest in fact a Galactocentric cosmic-ray 
emissivity gradient that is much smaller than that of the distribution of supernova remnants
or pulsars, which has been taken as evidence for large-scale cosmic-ray diffusion in our Galaxy
\citep{bloemen84a, bloemen84b, bloemen86, strong88, strong96}.
In addition, the low average gas density that is encountered by cosmic rays as determined
from cosmic-ray isotope ratios indicates that the particles spend a considerable fraction 
of their lifetimes outside the plane of the Galaxy, suggesting a cosmic-ray halo that 
extends to scale heights of 4-6 kpc above the galactic plane \citep{moskalenko04}.
On the other hand, \citet{fichtel84} have shown that the gamma-ray data are also
compatible with a tight coupling of cosmic rays to the matter, and \citet{strong04b}
suggest that neglecting a possible strong Galactocentric gradient in the 
CO-to-H$_2$ conversion factor has led in the past to an underestimation of the
Galactic cosmic-ray gradient.
Thus, observations may also be compatible with a short cosmic-ray diffusion length
in our own Galaxy.

Galactic studies are however hampered by our position within the Galactic plane,
which does not allow for an unambiguous correlation of the gamma-ray emissivity
with Galactic star forming regions, making estimates of cosmic-ray diffusion rather
uncertain.
Our observations of the LMC provide for the first time a much clearer picture about
the link between possible cosmic-ray acceleration sites and gamma-ray emission,
and suggest that both are tightly connected.
30~Dor would thus be the most powerful cosmic-ray acceleration site in the LMC,
which would be no surprise as it is one of the most active starburst regions in the entire
Local Group of galaxies \citep{pineda09}.

\subsection{Gamma rays from cosmic-ray interactions}
\label{sec:emissivity}

The gamma-ray emission from cosmic-ray interactions with the interstellar medium and 
radiation field is expected to be intrinsically diffuse in nature.
Indeed, our analysis suggests that the gamma-ray emission from the LMC does not originate
in a small number of individual point sources.
By adding successively point sources to our model (cf. section \ref{sec:geomodels})
we formally detected 5 significant point sources (model \psmod), but we found a model with 
less components that results in a higher TS value when we allow for the source components
to be extended (model \gmod).
It is thus more likely that the LMC emission is indeed diffuse in nature, or alternatively,
composed of a large number of unresolved and faint sources that can not be
detected individually by {\em Fermi}/LAT.

From here on we thus make the assumption that the gamma-ray
emission from the LMC indeed originates from cosmic-ray interactions with the interstellar 
gas and the stellar radiation field, and we discuss the implications of this assumption 
in light of our observations.
In particular we test the 2 extreme hypotheses that
(1) all gamma-ray emission from the LMC is attributed to diffuse emission from cosmic-ray
interactions, and
(2) only the emission from component G1 in the \gmod\ model arises from cosmic-ray
interactions, while the gamma rays from 30~Dor (or component G2 in the  \gmod\ model)
originate from other sources.
Hypothesis 2 is mainly motivated by the particular nature of 30~Dor, and we want to examine
how our conclusions are altered when we exclude this source from our considerations.
Note that in this case, only $70\%$ of the integrated LMC flux is attributed to cosmic-ray
interactions while for hypothesis 1 we assume that $100\%$ of the LMC flux is due
to cosmic rays.

Using these two hypotheses (which we call from here on \hone\ and \htwo) we computed the 
average integrated $>100$~MeV gamma-ray emissivity per hydrogen atom of the LMC 
using
\begin{equation}
q_{\gamma} = \frac{\Phi_{\gamma}}{\int N_{\rm H}~{\rm d}\Omega}
\label{eq:emissivity}
\end{equation}
where 
$\Phi_{\gamma}$ is the integrated $>100$~MeV photon flux from the LMC under
\hone\ or \htwo, and $\int N_{\rm H}~{\rm d}\Omega$ is the spatially integrated hydrogen 
column density of the LMC.
From the \hi\ map of \citet{kim05} one obtains
$\int N_{\rm H}~{\rm d}\Omega = 2.39 \times 10^{19}$~H-atom~cm$^{-2}$~sr
for atomic hydrogen, assuming optically thin emission.
Including molecular hydrogen increases the value by $5\%$ - $10\%$ and applying
optical depth corrections for 125~K and 60~K further increases the integral
by $12\%$ and $42\%$, respectively.
Evidence for some hidden hydrogen (or the presence of substantial optically thick
emission) comes from an analysis of infrared data obtained with {\em Spitzer},
suggesting an additional component that follows the distribution of the \hi\ seen
in the radio surveys, and that doubles the total gas mass \citep{bernard08}.
Thus, $\int N_{\rm H}~{\rm d}\Omega$ could be as large as 
$4.8 \times 10^{19}$~H-atom~cm$^{-2}$~sr
and we adopted
$\int N_{\rm H}~{\rm d}\Omega = (3.6 \pm 1.2) \times 10^{19}$~H-atom~cm$^{-2}$~sr
as a mean value and included the uncertainty of the total hydrogen mass of the LMC
in our computation.
The resulting emissivities are given in Table~\ref{tab:crresults}.

\begin{table}[!t]
\footnotesize
\caption{
Results of the gamma-ray emissivity analysis of the LMC for hypotheses \hone\ and \htwo.}
\label{tab:crresults}
\begin{center}
\begin{tabular}{lrr}
\hline\hline
Parameter & \multicolumn{1}{c}{\hone} & \multicolumn{1}{c}{\htwo} \\
\hline
$\Phi_{\gamma}$ & $25.0\pm1.7\pm4.1$ & $15.2\pm1.9\pm3.4$ \\
$q_{\gamma}$ & $6.9 \pm 2.4 \pm 2.6$ & $4.2 \pm 1.5 \pm 1.7$ \\
$r_c$ & $ 0.31 \pm  0.01 \pm  0.02~^{+0.13}_{-0.07}$ & $0.21 \pm 0.01 \pm  0.02~^{+0.09}_{-0.05}$ \\
\hline
\end{tabular}
\end{center}
Note to the table:
Given are the integrated $>100$~MeV photon flux $\Phi_{\gamma}$ in units of $10^{-8}$ \funit,
the average $>100$~MeV gamma-ray emissivity $q_{\gamma}$ in units of $10^{-27}$ \qunit,
and the cosmic-ray enhancement factor $r_c$.
Quoted errors are statistic and systematic.
For $r_c$ we quote in addition the (asymmetric) systematic error related to the
uncertainty in the total LMC gas mass.
\end{table}

\begin{figure}[!t]
\centering
\includegraphics[width=9.1cm]{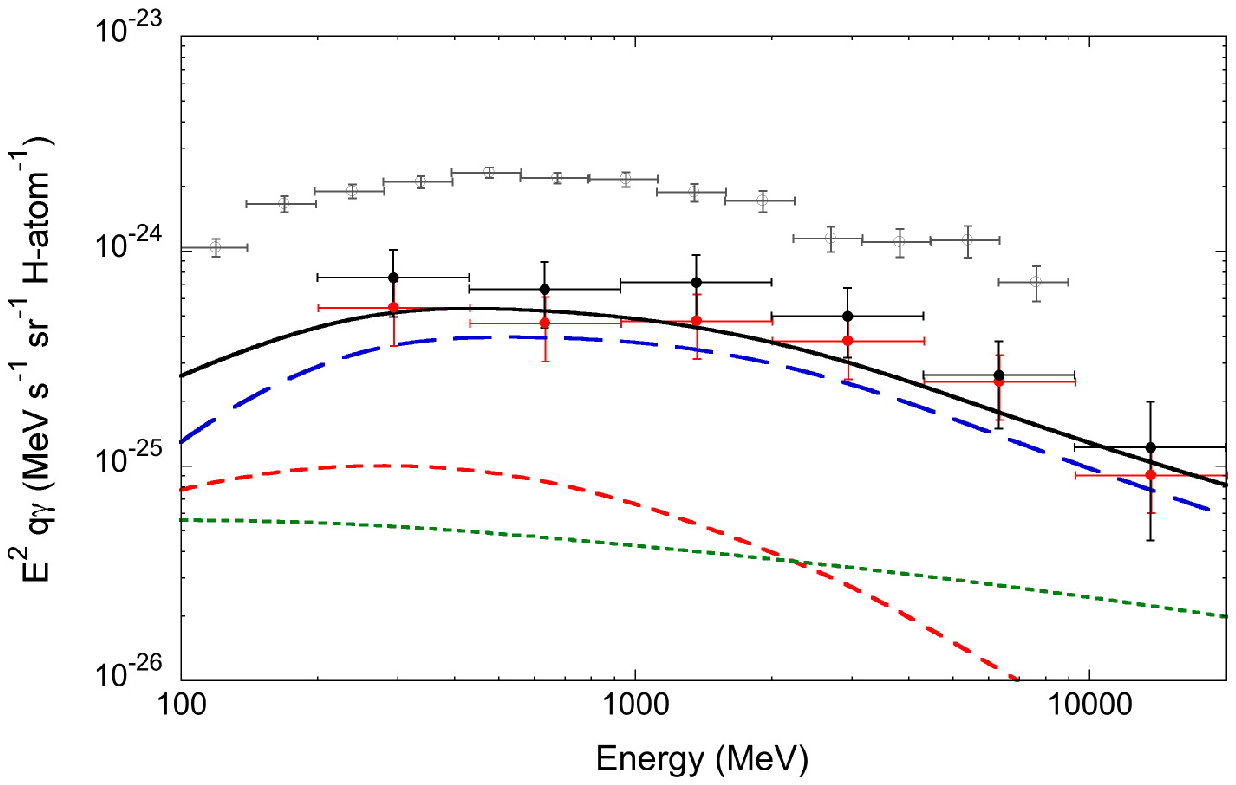}
\caption{
Differential average gamma-ray emissivity for \hone\ (black dots) and \htwo\ (red dots) 
compared to that of the Galactic local interstellar medium determined by 
\citet{abdo09c} (grey data points).
The error bars for \hone\ and \htwo\ include statistical and systematic errors and
the uncertainties in the total gas mass.
The solid line shows the predicted gamma-ray emissivity computed in the framework
of a one-zone model for \htwo\ (see text).
The other lines show the contributions of 
$\pi^0$-decay (long dashed),
Bremsstrahlung (short dashed), and
inverse Compton emission (dotted).
\label{fig:emissivityspectrum}
}
\end{figure}

Depending on the hypothesis we make, the average gamma-ray emissivity of the LMC 
is between $\sim2$ to $\sim4$ times smaller than the $>100$~MeV emissivity of
$q_{\gamma} = (1.63 \pm 0.05) \times 10^{-26}$ \qunit\
that has been determined by {\em Fermi}/LAT for the local interstellar medium of our
own Galaxy \citep{abdo09c}.
We illustrate this difference in Fig.~\ref{fig:emissivityspectrum}, where we compare the
differential gamma-ray emissivity of the local interstellar medium
\citep[cf.~Fig.~5 of][]{abdo09c}
to that of the LMC.
Following Eq.~(\ref{eq:emissivity}), we derived the differential gamma-ray emissivity by dividing 
our spectra by 
$\int N_{\rm H}~{\rm d}\Omega = (3.6 \pm 1.2) \times 10^{19}$~H-atom~cm$^{-2}$~sr.
For \hone\ we used the spectrum derived using the \hii\ gas map for the total emission from
the LMC (black dots), 
while for \htwo\ we used the LMC disk spectrum obtained using component G1 of
model \gmod\ (red dots)

We compared the differential gamma-ray emissivities to a one-zone model of 
cosmic-ray interactions with the interstellar medium that takes into account $\pi^0$ 
decay following proton-proton interactions, Bremsstrahlung from
cosmic-ray electrons and inverse Compton scattering of cosmic-ray electrons on LMC
optical and infrared photons and cosmic microwave background photons.
We calculated the $\pi^0$ production by proton-proton interactions following the 
prescription of \citet{kamae06}.
We used the cosmic-ray proton, electron and positron spectra presented in \citet{abdo09c} 
for the local Galactic environment.
The scaling factor ($r_c$) of our model with respect to the data, which we refer to as the cosmic-ray
enhancement factor, is thus a direct measure of the average LMC cosmic-ray density with respect 
to that in the vicinity of the Earth.
The $\pi^0$ and Bremsstrahlung emissivities were calculated by assuming the average
LMC metallicity to be $Z=0.4$ \Zsol\ \citep{westerlund97}.
We applied a corresponding nuclear enhancement factor of $\epsilon_{\rm M}=1.75$ to the
$\pi^0$ emissivities \citep{mori09} and took the total hydrogen mass of the LMC to be
$7.2 \times 10^8$ \Msol.\footnote{
  For a distance of 50 kpc to the LMC
  $\int N_{\rm H}~{\rm d}\Omega = (3.6 \pm 1.2) \times 10^{19}$~H-atom~cm$^{-2}$~sr
  corresponds to a total \hi\ mass of $(7.2 \pm 2.4) \times 10^8$ \Msol.}
The inverse Compton component was calculated using the method described by 
\citet{blumenthal70} using the optical and infrared interstellar radiation fields from 
\citet{porter08} that we rescaled according to the stellar luminosity density and the 
observed infrared emission \citep{bernard08}, respectively.

Fitting this spectral model to our data using a binned maximum likelihood analysis gave an 
average cosmic-ray enhancement factor of 
$r_c = 0.31 \pm 0.01$ for \hone, and of
$r_c = 0.21 \pm 0.01$ for \htwo\ (cf.~Table \ref{tab:crresults}).
Systematic errors due to uncertainties in the effective area of the instrument amount to
$\pm0.02$.
An additional systematic error of $-23\%$ to $+42\%$ comes from the uncertainty in the 
total gas mass of the LMC, which largely dominates the statistical and systematic
measurements errors.

\begin{figure*}[!th]
\centering
\includegraphics[width=9.1cm]{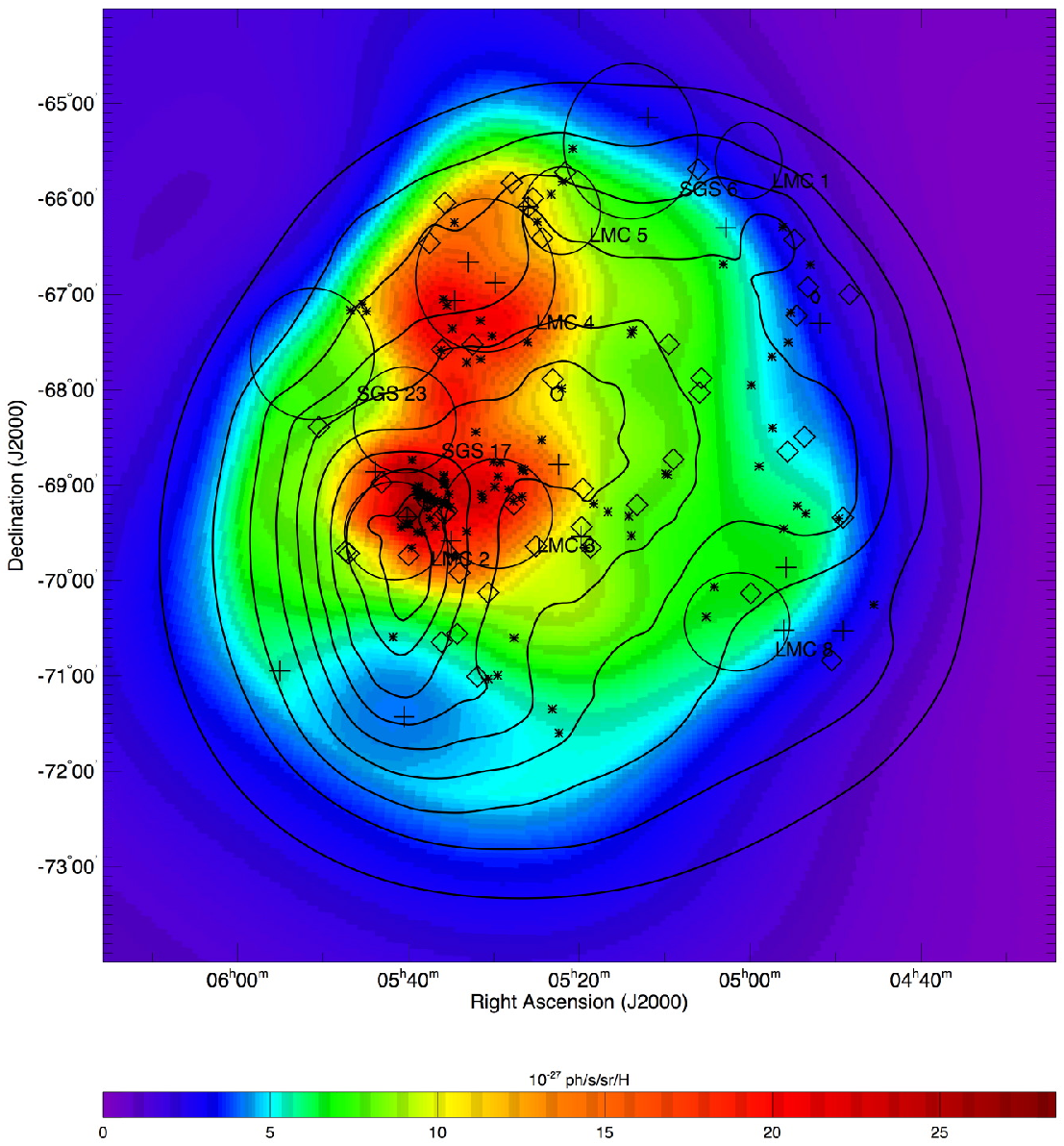}
\includegraphics[width=9.1cm]{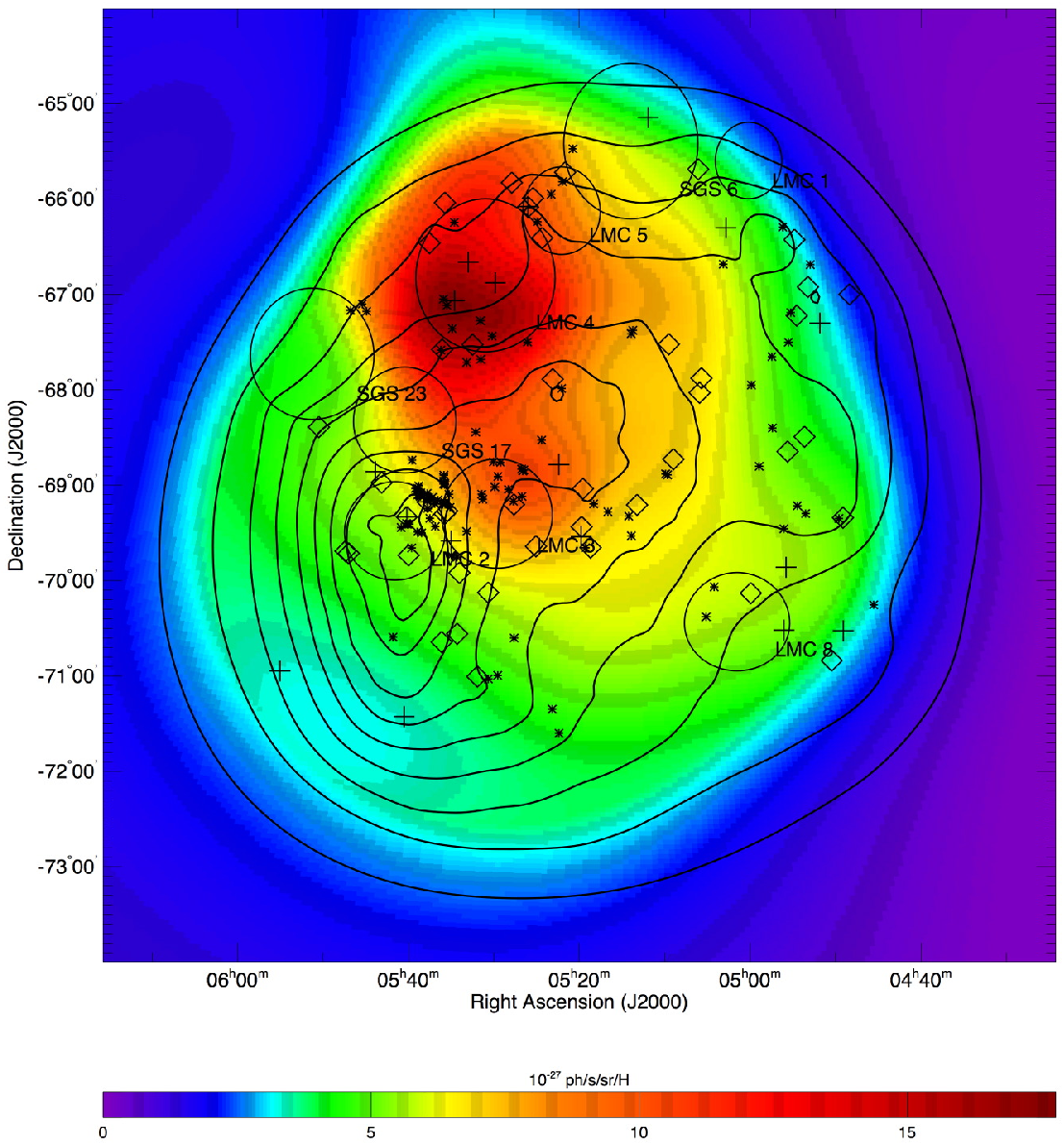}
\caption{
Integrated $>100$~MeV emissivity maps of the LMC in units of
$10^{-27}$~ph~s$^{-1}$~sr$^{-1}$~H-atom$^{-1}$ 
for \hone\ (left panel) and \htwo\ (right panel).
An adaptive smoothing with a signal-to-noise ratio of 5 has been applied to reduce statistical
fluctuations.
Note that the colour scales are not the same for the two images.
As contours we show $N({\rm H})$ column densities convolved with the LAT point spread 
function that are linearly spaced from $10\%$ - $90\%$ of the maximum in steps of $10\%$.
Symbols indicate the locations of 
pulsars (pluses) from ATNF catalogue version 1.36 \citep{manchester05},
supernova remnants (diamonds) from Rosa Williams web page 
http://www.astro.illinois.edu/projects/atlas/index.html,
Wolf-Rayet stars (stars) from the fourth catalogue of \citet{breysacher99},
and supergiant shells (circles) from \citet{staveleysmith03}.
\label{fig:emissivitymap}
}
\end{figure*}

\citet{fichtel91} estimated the cosmic-ray enhancement factor to be $r_c\sim1$ by comparing 
the synchrotron radiation of the LMC to that of the Milky Way (cf.~their Fig.~3).
We also computed the synchrotron emission in our model under the assumption that the
cosmic-ray proton to electron ratio in the LMC and the electron spectrum are the same 
than those measured in the local interstellar medium \citep{abdo09c}.
We further took the thickness of the emitting region to be 2 kpc and assumed a magnetic
field of 5 $\mu$G \citep{pohl93}.
Our predicted synchrotron flux is in good agreement with the observed non-thermal flux
quoted by \citet{klein89}.
\citet{fichtel91}, however, assumed a higher non-thermal flux for the LMC compared to the
observations of \citet{klein89}, which may explain why their $r_c$ is in excess of our
model.

\citet{pavlidou01} followed another approach and predicted $r_c=0.14$ from a comparison
of the supernova rate in the LMC to that in our Galaxy, yet they considered also higher values
due to the large uncertainties that are inherent to the estimates of supernova rates.
Their estimate relies further on the assumption that the cosmic-ray containment in the LMC is 
similar to that in our own Galaxy, which is not necessarily fulfilled since cosmic rays may escape
fairly rapidly into intergalactic space because of the small size of the LMC
\citep{ginzburg85}.
Our value, which is slightly above the estimate of \citet{pavlidou01}, suggests however that
cosmic-ray escape from the LMC is not more important than for our own Galaxy.
This is in line with the relatively small proton diffusion length that we inferred from the
compactness of the 30~Dor source (cf.~section~\ref{sec:30dor}).

\subsection{The sites of cosmic-ray acceleration in the LMC}

To reveal the sites of cosmic-ray acceleration in the LMC we mapped the cosmic-ray density 
variations in the galaxy by computing the gamma-ray emissivity $q_{\gamma}$ as function 
of position.
We did this by generating a background subtracted counts map for \hone, and a background
and G2 model subtracted counts map for \htwo\ that we divided by the $N({\rm H})$ map after
convolution of the latter with the LAT instrumental response function.
We normalised $N({\rm H})$ to a total LMC hydrogen mass of 
$7.2 \times 10^8$ \Msol\
that takes into account the possible presence of dark gas that is not seen in radio surveys
of \hi\ (see section \ref{sec:emissivity}).
We adaptively smoothed \citep{ebeling06} the counts maps and used the resulting smoothing
kernel distribution to smooth also the convolved $N({\rm H})$ map before the division to 
reveal significant structures at all possible scales, while suppressing the noise that arises 
from the limited photon counting statistics.

The resulting emissivity maps are shown in Fig.~\ref{fig:emissivitymap}.
We superimpose on the images the interstellar gas distribution, as traced by $N({\rm H})$, 
convolved with the LAT instrumental response function, and also show the locations of 
potential particle acceleration sites, such as pulsars, supernova remnants, Wolf-Rayet stars 
and supergiant shells.

Figure~\ref{fig:emissivitymap} reveals that the cosmic-ray density varies considerably
over the disk of the LMC.
The gamma-ray emissivity is highest in 30~Dor and the northern part of the galaxy, while 
the southern part and in particular the dense ridge of gas south of 30~Dor seems basically
devoid of cosmic rays.
The large variations in $q_{\gamma}$ confirm our earlier findings 
(cf.~section \ref{sec:templatefitting})
that the gamma-ray emission correlates little with the gas density in the LMC, and this is 
irrespective of whether we consider 30~Dor to be part of the emission
(left panel versus right panel).
The most striking variation is found along the gas ridge that runs over $\sim3\deg$ along 
$\ra\sim05^{\rm h}40^{\rm m}$ which coincides with the most prominent region of
$^{12}$CO emission tracing giant molecular clouds in the LMC \citep{fukui99}.
Roughly $20\%$ of the total gas mass in the LMC is confined into this ridge \citep{luks92}, 
and if the cosmic-ray density were uniform over the LMC (or at least over the ridge area) the 
entire ridge should be a source of high-energy gamma rays.
This, however, is obviously not the case.
Cosmic rays apparently do not penetrate into the southern part of the ridge, which is an
additional argument in favour of a short GeV cosmic-ray proton diffusion length in the 
LMC.

Figure~\ref{fig:emissivitymap} suggests further that the cosmic-ray density correlates
with massive star forming tracers, and in particular Wolf-Rayet stars and supergiant 
shells.
This correlation is particularly striking if we assume that the emission from 30~Dor is 
originating in the same physical process as the rest of the LMC emission (\hone).
If we exclude 30~Dor from the comparison (\htwo) the correlation becomes less 
striking, yet still, most of the cosmic rays are found in regions that are rich in Wolf-Rayet 
stars and that are located near supergiant shells.
This finding is corroborated by the good fit of the \hii\ gas map (which is probably the most
direct tracer of massive star forming regions within a galaxy), even if the 30~Dor emission
is discarded from the analysis (cf.~section~\ref{sec:templatefitting}).

Thus, the gamma-ray emissivity maps of the LMC support the idea that cosmic rays
are accelerated in massive star forming regions as a result of the large amounts of kinetic
energy that are input by the stellar winds and supernova explosions of massive stars into
the interstellar medium.
Our data reveal a relatively tight confinement of the gamma-ray emission
to star forming regions, which suggests a relatively short diffusion length for GeV
protons.

\section{Conclusions}
\label{sec:conclusion}

Observations of the LMC by {\em Fermi}/LAT have for the first time provided a detailed
map of high-energy gamma-ray emission from that galaxy.
Our analysis revealed the massive star forming region 30~Doradus as bright source of
gamma-ray emission in the LMC in addition to fainter emission regions found in the northern
part of the galaxy.
So far, we could not identify any point source contributions to the emission from the
30~Dor region, and in particular we did not significantly detect pulsations from 
the energetic Crab-like pulsars \psra\ and \psrb, although the lightcurve of the former
indicates variability at the $2.4\sigma$ level.
Cosmic-ray interactions with the interstellar medium and radiation field seem thus the
most plausible origin for the observed gamma-ray emission from 30~Dor.

The gamma-ray emission from the LMC shows very little correlation
with gas density.
A much better correlation is seen between gamma-ray emission and massive star
forming regions, as traced by the ionizing gas, Wolf-Rayet stars and supergiant shells,
and we take this as evidence for cosmic-ray acceleration in these regions.
The close confinement of gamma-ray emission to star forming regions implies
a relatively small GeV cosmic-ray protons diffusion length.

Continuing observations of the LMC with {\em Fermi}/LAT in the upcoming
years will provide the photon statistics to learn more about the origin of the gamma-ray
emission from that galaxy.
Better statistics will help in identifying more individual emission components and may help
to separate true point sources from the more diffuse emission that we expect from
cosmic-ray interactions.
In particular, 1-2 year's of additional exposure should allow the question of the
contribution of \psra\ to the gamma-ray emission from 30~Dor to be answered,
and should also allow clearly establishing the pulsar wind nebulae contributions at 
higher energies.

\begin{acknowledgements}
The \textit{Fermi} LAT Collaboration acknowledges generous ongoing support
from a number of agencies and institutes that have supported both the
development and the operation of the LAT as well as scientific data analysis.
These include the National Aeronautics and Space Administration and the 
Department of Energy in the United States, the Commissariat \`a l'Energie Atomique
and the Centre National de la Recherche Scientifique / Institut National de Physique
Nucl\'eaire et de Physique des Particules in France, the Agenzia Spaziale Italiana
and the Istituto Nazionale di Fisica Nucleare in Italy, the Ministry of Education,
Culture, Sports, Science and Technology (MEXT), High Energy Accelerator Research
Organization (KEK) and Japan Aerospace Exploration Agency (JAXA) in Japan, and
the K.~A.~Wallenberg Foundation, the Swedish Research Council and the
Swedish National Space Board in Sweden.

Additional support for science analysis during the operations phase is gratefully
acknowledged from the Istituto Nazionale di Astrofisica in Italy and the and the Centre National d'\'Etudes Spatiales in France.
\end{acknowledgements}


\end{document}